\documentclass[10pt,journal,twocolumn,romanappendices]{IEEEtran}
\IEEEoverridecommandlockouts
\pdfoutput=1
\usepackage{algorithm,algorithmic,amsmath,amssymb,amsthm,array,bbm,cite,color,comment,graphicx,microtype,setspace,url,caption}
\usepackage[USenglish]{babel}
\usepackage{hyperref}
\usepackage[utf8]{inputenc}
\usepackage[T1]{fontenc}
\usepackage{enumitem}
\usepackage{graphicx} 
\usepackage{subcaption} 
\usepackage{xcolor}
\usepackage{soul}

\setitemize{itemsep=0mm,topsep=0mm,parsep=0pt,partopsep=0pt}

\usepackage{tikz}
\usetikzlibrary{arrows,backgrounds,calc,intersections,decorations.pathreplacing,positioning,shapes}

\makeatletter
\newcommand\subparagraph{%
  \@startsection{subparagraph}{5}
  {\parindent}
  {3.25ex \@plus 1ex \@minus .2ex}
  {-1em}
  {\normalfont\normalsize\bfseries}}
\makeatother

\usepackage{titlesec}
\let\subparagraph\relax
\titlespacing{\section}{3pt}{4pt plus 2pt minus 1pt}{3pt plus 2pt minus 1pt}
\titlespacing{\subsection}{3pt}{3pt plus 1pt minus 0pt}{2pt plus 1pt minus 0pt}
\usepackage{todonotes}
\usepackage[font=small]{caption}
\usepackage{comment}
\usepackage{epstopdf}
\usepackage{pgfplots}
\usepgfplotslibrary{groupplots}
\pgfplotsset{compat=1.17}
\usepackage{algorithm,algorithmic,amsmath,amssymb,amsthm,array,bbm,cite,color,comment,graphicx,microtype,setspace,url,caption}

\usepackage[USenglish]{babel}
\usepackage{xcolor}
\usepackage{soul}
\usepackage{hyperref}
\usepackage[utf8]{inputenc}
\usepackage[T1]{fontenc}
\usepackage{enumitem}
\usepackage{graphicx} 
\usepackage{subcaption} 
\usepackage{booktabs} 
\usepackage{booktabs,multirow,siunitx,caption}
\usepackage{tabularx,array,amsmath}
\usepackage{tabularx,array,makecell,amsmath}
\usepackage{tabularx,array,makecell,amsmath}
\usepackage[T1]{fontenc}
\usepackage[utf8]{inputenc}
\usepackage{graphicx}
\DeclareGraphicsExtensions{.pdf,.png,.jpg,.eps}
\usepackage{graphicx}
\usepackage{epstopdf}

\setitemize{itemsep=0mm,topsep=0mm,parsep=0pt,partopsep=0pt}

\usepackage{tikz}
\usetikzlibrary{arrows,backgrounds,calc,intersections,decorations.pathreplacing,positioning,shapes}

\makeatletter
\newcommand\subparagraph{%
  \@startsection{subparagraph}{5}
  {\parindent}
  {3.25ex \@plus 1ex \@minus .2ex}
  {-1em}
  {\normalfont\normalsize\bfseries}}
\makeatother

\usepackage{titlesec}
\let\subparagraph\relax
\titlespacing{\section}{3pt}{4pt plus 2pt minus 1pt}{3pt plus 2pt minus 1pt}
\titlespacing{\subsection}{3pt}{3pt plus 1pt minus 0pt}{2pt plus 1pt minus 0pt}
\usepackage{todonotes}
\usepackage[font=small]{caption}
\usepackage{comment}
\usepackage{epstopdf}
\usepackage{pgfplots}
\usepgfplotslibrary{groupplots}
\pgfplotsset{compat=1.17}

\let\originalcaron\v

\renewcommand{\v}{\mathbf{v}}

\newcommand{\x}{\mathbf{x}}

\newcommand{\setI}{\mathcal{I}}

\newcommand{\setM}{\mathcal{M}}

\newcommand{\setT}{\mathcal{T}}
\newcommand{\setU}{\mathcal{U}}

\newcommand{\diag}{\mathrm{diag}}

\newcommand{\Exp}{\mathbb{E}}
\newcommand{\herm}{\mathrm{H}}
\renewcommand{\Im}{\mathrm{Im}}

\renewcommand{\Re}{\mathrm{Re}}

\newcommand{\tran}{\mathrm{T}}

\title{Near-Field MIMO LoS Channel Recovery Under User Antenna Asymmetry}
\author{Shima Eslami,~\IEEEmembership{Graduate Student Member,~IEEE}, Jarkko Kaleva,~\IEEEmembership{Member,~IEEE}, \\ and Antti Tölli,~\IEEEmembership{Senior Member,~IEEE}
\thanks{The authors are with the Centre for Wireless Communications, University of Oulu, Finland {(e-mail: \{shima.eslami, jarkko.kaleva, antti.tolli\}@oulu.fi )}. Author Jarkko Kaleva is also with Elisa Oyj {(email: jarkko.kaleva@elisa.fi)}. This work was supported by the Research Council of Finland under Projects 368243 (INCOMMS) and 346208 (6G Flagship).
}}
\begin{document}

\maketitle

\begin{abstract}
Near-field (N-F) line-of-sight (LoS) MIMO channels admit a compact representation through a small number of geometric parameters describing the relative transceiver geometry, enabling low-overhead channel acquisition without estimating the full channel matrix. This paper considers downlink (DL) channel acquisition under practical asymmetric user-equipment (UE) antenna capabilities with multiple DL receive antennas, but only a single active uplink (UL) transmit chain. 
In this setting, UL observations across the large BS aperture enable accurate estimation of the relative location of the UL-active UE antenna but are insufficient to determine the orientation of the entire UE array.
We propose a two-stage geometry-based N-F LoS MIMO channel acquisition framework that distributes the estimation of the channel-defining geometric parameters between the BS and UE. First, a single UL pilot sequence enables the BS to estimate two BS-side reference angles that parameterize the relative location of the UL-active UE antenna. The BS conveys these estimates to the UE and transmits two DL pilot sequences, from which the UE estimates the remaining array-orientation parameter.
We derive Cramér–Rao lower bounds (CRLBs) for the first-stage parameters and a first-order error-covariance approximation for the UE orientation that explicitly captures the propagation of first-stage estimation uncertainty. Numerical results characterize the resulting error propagation and demonstrate accurate estimation of the geometric parameters and reconstruction of the LoS MIMO channel using one UL and two DL pilot sequences.
\end{abstract}
\bstctlcite{IEEEexample:BSTcontrol}
\section{Introduction}
\label{section:1}
Near-field (N-F) communication is an emerging transmission regime for future 6G wireless systems~\cite{A_Tutorial}, particularly at higher carrier frequencies and with electrically large antenna apertures~\cite{Comprehensive_Survey,channel_est_survey}.
Unlike far-field (F-F) propagation, characterized by planar wavefronts and angular channel information, N-F propagation exhibits spherical wavefronts that depend on both angle and distance,
enabling spatial focusing toward specific locations and finer spatial resolution~\cite{channel_est_survey, Emil1, NF_Engineering }. These capabilities can improve spatial multiplexing, spectral efficiency, achievable rates, and localization accuracy~\cite{A_Tutorial, Polar-Domain-Codebooks, Tutorial_NF_driven},
making N-F communication attractive for
applications such as massive connectivity, high-accuracy localization and sensing, integrated sensing and communication, 
and enhanced physical-layer security~\cite{Comprehensive_Survey,future_direction, Tutorial_NF_driven}. These benefits require accurate channel state information (CSI) at the base station (BS)~\cite{future_direction, Research_Advances}. 

The spherical-wavefront nature of N-F propagation introduces additional challenges for channel acquisition compared with the F-F case~\cite{Polar-Domain-Codebooks,sensing_CE, Research_Advances}.
In particular, conventional F-F acquisition methods based primarily on angular information may fail to capture the distance-dependent phase variations across large antenna apertures~\cite{channel_est_survey,future_direction, Research_Advances}, while direct estimation of the high-dimensional channel coefficients can incur substantial pilot and signaling overhead~\cite{sensing_CE,Research_Advances}.
Existing studies have shown that N-F channels exhibit exploitable structure, including sparsity in the joint angle–distance (polar) domain~\cite{Polar-Domain-Codebooks, NF_Deep_Generative_Model, Localization, Dai_mixedLoS, NF_channel_est_UPA} and, for line-of-sight (LoS) MIMO channels, a low-dimensional dependence on the deterministic transceiver geometry~\cite{Dai_mixedLoS, Nitin,myICC, NF_LoS_MIMIO_3D, Shima}. These properties enable structured channel-acquisition methods that avoid direct estimation of all channel coefficients, either by recovering a sparse set of polar-domain coefficients or path parameters~\cite{Polar-Domain-Codebooks, NF_Deep_Generative_Model, Dai_mixedLoS, NF_channel_est_UPA}, or by estimating a small set of geometric parameters from which the channel can be reconstructed~\cite{Dai_mixedLoS, Localization, Nitin,myICC, NF_LoS_MIMIO_3D, Shima}.
Several works exploit polar-domain sparsity and/or angle–distance parameterizations of N-F MISO/SIMO channels for CSI acquisition.
The authors of~\cite{Polar-Domain-Codebooks} investigate the design of polar-domain codebooks for channel estimation and beam training, aiming to reduce codebook dimensionality and acquisition overhead.
%
%
In~\cite{NF_Deep_Generative_Model}, a compressed sensing (CS)-based channel estimate is subsequently refined using a generative model to mitigate the impact of discrete angle–distance sampling and improve performance under limited communication overhead.
%
In~\cite{Localization}, a sparse-recovery approach estimates the angular and distance parameters of each channel path using a maximum-likelihood criterion, followed by damped Newton-based refinement.
%
%
%
%
In \cite{Dai_mixedLoS}, polar-domain sparsity is exploited for CS-based estimation of the non-LoS (NLoS) MIMO channel by modeling each path as a cascade of a TX-to-scatterer MISO link and a scatterer-to-RX SIMO link.
In \cite{NF_channel_est_UPA}, a modified NLoS channel model with joint angular–distance parameterization is proposed, followed by N-F channel-parameter estimation through coarse initialization and subsequent 
refinement.

A second line of work focuses on LoS MIMO CSI acquisition by directly estimating a small set of transceiver-geometry parameters.
In \cite{Nitin}, the 
N-F MIMO channel is partitioned into locally F-F subchannels, whose structural dependencies are exploited for low-overhead channel estimation and beam alignment.
%
%
%
Reference \cite{myICC} estimates two reference TX-antenna locations from their SIMO-channel AoAs and exploits the geometric relations among the TX antennas to infer the remaining antenna locations and reconstruct the full N-F LoS MIMO channel.
In \cite{Dai_mixedLoS} the N-F LoS MIMO channel is parameterized by three geometric parameters, which are jointly estimated through coarse search and gradient-based refinement.
%
More recently, \cite{NF_LoS_MIMIO_3D} characterizes the LoS MIMO channel between TX and RX arrays in 3D space using five geometric parameters and estimates them through a nonuniform search followed by gradient-based refinement.
The above approaches generally rely on user-specific uplink (UL) pilot observations, with the associated overhead increasing with the number of served user equipment (UEs) and the number of antennas per UE.
%
As an alternative solution,~\cite{Shima} proposes a DL pilot-based LoS CSI acquisition strategy in which a fixed, UE-independent set of DL pilots is transmitted from a few selected BS antennas. Each UE estimates its geometric parameters (i.e., two AoDs and relative UE-BS orientation) from its own DL observations obtained from the common pilot transmissions and feeds them back to the BS for channel reconstruction.
%
%
%
%

Focusing on N-F channel-acquisition methods based on geometric parameter estimation, a common principle is to first obtain channel observations through a limited number of appropriately designed pilot transmissions, then estimate a compact set of physical geometric parameters whose dimension does not scale directly with the number of antennas, and finally use the estimated parameters locally~\cite{Dai_mixedLoS, Localization, Nitin,myICC, NF_LoS_MIMIO_3D} or feed them back for CSI reconstruction~\cite{Shima}.
Such parameterized representations can reduce pilot and parameter-feedback overhead and improve robustness by restricting estimation to physically feasible channel geometries. Moreover, since the underlying geometric parameters are carrier-frequency invariant, geometry-based CSI acquisition can be applied to both time-division duplex (TDD) and frequency-division duplex (FDD) systems.
For example,~\cite{FDD-Massive-MIMO} exploits frequency-invariant parameters obtained from UL pilots to reconstruct the F-F DL channel in FDD massive MIMO, while~\cite{Limited-Feedback} uses limited angle and range feedback for N-F FDD extremely large-scale MIMO.
%

In practical UE implementations, the number of antennas available for UL transmission is often smaller than that available for DL reception due to hardware-complexity and power-consumption constraints~\cite{ARDIRef}. Among the aforementioned works, \cite{myICC} can accommodate such asymmetry, since the UL pilots are transmitted by only a subset of the UE antennas.
%
%
Similarly, \cite{ARDIRef} considers UL pilot transmission from a subset of UE antennas and reconstructs the full DL channel by inferring the propagation parameters of the non-transmitting antennas from those estimated for the transmitting antennas and a priori knowledge of the antenna layout.

A distinct challenge, however, arises when only a single UE antenna is active for UL transmission while multiple antennas are available for DL reception.
%
In this case, the UL observations collected across the BS aperture enable localization only of the single UL-active UE antenna and are therefore insufficient to determine the UE-array orientation.
The DL pilot-based approach in~\cite{Shima} is applicable to this setting, as it does not rely on multi-antenna UL pilot transmission and enables LoS CSI acquisition at the BS with limited pilot and geometric-feedback overhead.
However, even a single UL-active antenna provides an opportunity to exploit the spatial observations across the large BS aperture and shift part of the geometric-parameter estimation to the BS.
Motivated by this observation, this paper proposes a hybrid UL/DL geometry-based N-F LoS
\footnote{We focus on LoS CSI acquisition; residual NLoS components, when present, can be handled using the LoS-CSI-aided compensation approach in~\cite{Shima}.}
MIMO CSI acquisition strategy for asymmetric UE transmit/receive capabilities, particularly when only a single UE antenna is available for UL transmission. Building on the same N-F LoS channel parametrization as in~\cite{Shima}, the approach exploits geometric information obtained and exchanged through UL and DL signaling to distribute the parameter-estimation task between the BS and UE according to the parameters observable at each side. This enables full LoS MIMO CSI reconstruction without requiring UL pilot transmission from all UE antennas and can also improve estimation quality by exploiting the large BS array for UL estimation of the reference-location parameters, depending on the available UL pilot power.
The main contributions of our work are summarized as:
\begin{itemize}    
%
\item 
%
The proposed procedure follows a two-stage UL/DL parameter-estimation process. In the first stage, a single UL pilot sequence from the active UE antenna enables the BS to estimate two BS-side reference angles that determine the relative location of this antenna. The BS then conveys these estimates to the UE and transmits two DL pilot sequences, from which the UE estimates the UE-array orientation parameter and feeds it back for LoS MIMO channel reconstruction.This split reduces the UE-side estimation from the full set of geometric parameters to only the UE-array orientation.
 \item 
 %
 We analyze the estimation errors across the two stages by deriving the CRLBs for the jointly estimated first-stage angles and a first-order covariance approximation for the UE-array orientation that accounts for uncertainty propagated through the geometric mapping from the first-stage estimates.
     %
     \item Numerical results characterize the estimation and channel-reconstruction performance across SNR regimes and quantify the impact of first-stage estimation errors and parameter quantization. They also show that propagated first-stage errors determine an error floor in UE-array orientation estimation and that, with sufficiently accurate UL geometry estimates, two DL pilot sequences achieve channel-reconstruction accuracy comparable to the considered DL-only geometry-based scheme.
\end{itemize}
The paper is structured as follows: Section~\ref{sec:SM} introduces the system and channel models. Section~\ref{sec:ch-est} presents the proposed two-stage UL/DL-assisted LoS CSI acquisition framework and the corresponding estimators. Section \ref{sec:CRLB} provides the first-order error analysis of both estimation stages. Section \ref{sec:num} presents the numerical results, and Section \ref{sec:CONC} concludes the paper.
 
 \textit{Notation:} For a matrix $\mathbf{X}$, $\mathbf{X}^\tran$, $\mathbf{X}^\herm$, and $\mathbf{X}^{-1}$ denote the transpose, conjugate transpose, and inverse, respectively. The entry in the $n^\text{th}$ row and $m^\text{th}$ column of $\mathbf{X}$ is $\mathbf{X}_{[n,m]}$, and the $m^\text{th}$ column is $\mathbf{X}_{[:,m]}$. The real and imaginary parts of $\mathbf{X}$ are $\Re(\mathbf{X})$ and $\Im(\mathbf{X})$, respectively.
$\diag({x}_1,\cdots,{x}_N)$ creates a diagonal matrix with elements of ${x}_1,\cdots,{x}_N$ on the diagonal. Horizontal and {vertical} concatenation of vectors $\mathbf{x}_n$ is written as $[\mathbf{x}_1,\cdots,\mathbf{x}_N]$ and {$[\mathbf{x}_n]_{n=1}^N$}. 
For a set $\setM$, $\setM(i)$ is its $i^\text{th}$ member, and $|\setM|$ is its cardinality. 
The Euclidean norm is $\| \cdot \|$.
The notation $\mathbf{F}(\x)$ or $\mathbf{F}(\x_{[1]},\cdots,\x_{[M]})$ indicates that each element of $\mathbf{F}$ is a function of components of $\x$.
\section{System Model}
\label{sec:SM}
%
%
\begin{figure}[t!]
    \centering   \includegraphics[scale=0.5]{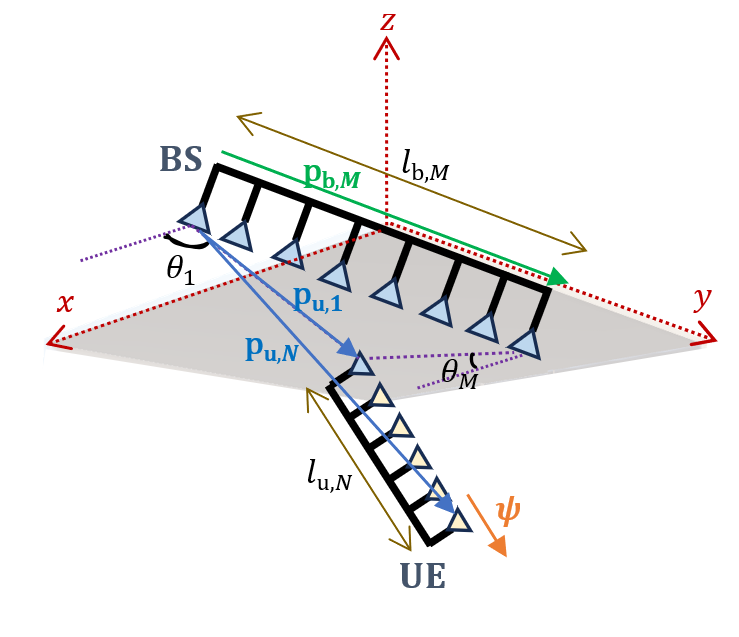} 
    \caption{{Geometric parameters of an N-F LoS MIMO channel.}}
    \label{fig:system_model1}
\end{figure}
We consider a DL MIMO system where the BS serves a UE in the N-F region. We assume that the linear antenna arrays at the BS and UE have a fully digital architecture and are equipped with $M$ and $N$ elements, respectively. 
Without loss of generality, we assume the first UE antenna is UL-active and is taken as the reference antenna, while all $N$ UE antennas are available for DL reception.
Focusing on the LoS channel component, the N-F MIMO channel $\mathbf{H}$ is modeled as~\cite{Indoor_mmw}
\begin{align} \label{eq:H_LoS}
\mathbf{H}_{[n,m]} = \frac{\lambda}{4\pi r_{n,m}}
\:\textit{exp}\:\bigg(\frac{-\textsf{j}2\pi r_{n,m}} {\lambda}\bigg),
\end{align}
where $\mathbf{H}_{[n,m]}$ is the $(n,m)^\text{th}$ entry of the channel matrix, 
$r_{n,m}$ is the distance between the $m^\text{th}$ BS antenna and $n^\text{th}$ UE antenna, and $\lambda$ is the carrier wavelength.
In general, the UL and DL carrier wavelengths may differ; accordingly, $\lambda=\lambda_\text{UL}$ for the UL channel and $\lambda=\lambda_\text{DL}$ for the DL channel. 
While the resulting UL and DL channel coefficients may differ, the underlying transceiver geometry (e.g., $r_{n,m}$) remains common to both links.
To characterize $\mathbf{H}$ using geometric parameters, we consider the setup in Fig.~\ref{fig:system_model1}, similarly to~\cite {Shima}, and define a 3D coordinate system
\footnote{This coordinate system is defined to determine the relative positions 
of the BS and UE antenna elements and does not necessarily coincide with the global 3D coordinate system.} 
in which the $xy$-plane contains the BS array and the reference UE antenna. The BS array is aligned with the $y$-axis 
and centered at the origin,
while the $z$-axis is perpendicular to the $xy$-plane. 
As shown in Fig.~\ref{fig:system_model1}, the location vector of the reference UE antenna relative to the first BS antenna, $\mathbf{p}_{\text{u},1}$, is determined by the intersection of the two BS-side reference angles $(\theta_1,\theta_M)$,  
referred to as AoDs, as
\begin{align} \label{eq:p_u,1}
\mathbf{p}_{\text{u},1}(\theta_1,\theta_M) = r_{1,1}(\theta_1,\theta_M) [\cos\theta_1, \sin\theta_1, 0 ]^\tran,
\end{align}
where $\theta_1$ and $\theta_M$ are the angle between the direction of the $x$-axis and $r_{1,1}$, and $r_{1,M}$, respectively. Moreover,
\begin{align} \label{eq:r11}
r_{1,1}(\theta_1,\theta_M) = l_{\text{b},M} \cos\theta_M \Big(\sin(\theta_M-\theta_1)\Big)^{-1} ,
\end{align}
where $l_{\text{b},M}$ is the length of the BS array. 
Accordingly, the location vector of the $n^\text{th}$ UE antenna can be determined from $\mathbf{p}_{\text{u},1}(\theta_1,\theta_M)$ and the UE orientation vector $\boldsymbol{\psi}$, as illustrated in Fig.~\ref{fig:system_model1}. We have 
\begin{align} \label{eq:p_u,n}
\mathbf{p}_{\text{u},n}(\boldsymbol{\psi},\theta_1,\theta_M) = l_{\text{u},n}\boldsymbol{\psi}+\mathbf{p}_{\text{u},1}(\theta_1,\theta_M),\quad \forall n\in \setI_N,
\end{align}
where $\setI_N=\{1,2,\cdots,N\}$ and $l_{\text{u},n}$ is the distance between the first and the $n^\text{th}$ UE antenna, which is assumed to be known for evaluating~\eqref{eq:p_u,n}. 
The orientation vector $\boldsymbol{\psi}$ in the 3D coordinate system has a unit norm and can be characterized by an azimuth ($\varphi$) and an elevation angle ($\omega$), i.e., 
\begin{equation}
\label{eq:psi_vector}
   \boldsymbol{\psi}=[\sin\omega\sin\varphi, \sin\omega\cos\varphi,\cos\omega]^\tran. 
\end{equation}
Let $l_{\text{b},m}, \forall m\in \setI_M$, denote the distance between the $m^\text{th}$ and first BS antennas. The location vector of the $m^\text{th}$ BS antenna relative to the first one is then given by $\mathbf{p}_{\text{b},m}=l_{\text{b},m}[0,1,0]^\tran$. Finally, $r_{n,m}$ is 
\begin{align} \label{eq:r_nm}
r_{n,m}(\boldsymbol{\psi},\theta_1,\theta_M) = \|\mathbf{p}_{\text{u},n}(\boldsymbol{\psi},\theta_1,\theta_M)-\mathbf{p}_{\text{b},m}\|.
\end{align}
Substituting \eqref{eq:r_nm} into \eqref{eq:H_LoS} shows that the N-F LoS MIMO channel is characterized by the two AoDs and the 3D relative orientation vector. Hence, given the known inter-antenna spacings 
$\{l_{\mathrm{b},m}\}_{m\in\mathcal{I}_M}$ and $\{l_{\mathrm{u},n}\}_{n\in\mathcal{I}_N}$, 
acquisition of $\mathbf{H}$ reduces to estimating the geometric parameters $(\theta_1,\theta_M,\boldsymbol{\psi})$. 
Moreover, since the relative transceiver geometry is fixed, the parameter set $(\theta_1,\theta_M,\boldsymbol{\psi})$ is frequency invariant, while the channel entries $\mathbf{H}_{[n,m]}$ in \eqref{eq:H_LoS} generally differ for $\lambda_\text{UL}$ and $\lambda_\text{DL}$. Thus, estimation of 
$(\theta_1,\theta_M,\boldsymbol{\psi})$ does not require complex-channel reciprocity and applies to both TDD and FDD operation.
%



\section{Proposed DL CSI acquisition under UE Antenna Asymmetry}
\label{sec:ch-est}
In this section, we describe the proposed two-stage acquisition of the N-F LoS MIMO channel parameters $(\theta_1,\theta_M,\boldsymbol{\psi})$ at the BS assuming asymmetric UL/DL antenna capabilities at the UE.
In the first stage, a single UL pilot from the reference UE antenna is observed across the BS aperture, enabling the BS to estimate $(\theta_1,\theta_M)$, which are then conveyed to the UE and used to determine the UE reference-antenna position $\mathbf{p}_{\text{u},1}$. 
In the second stage, two DL pilot sequences enable the UE to estimate the remaining parameter, i.e., the UE orientation $\boldsymbol{\psi}$, which is then fed back to the BS. The BS then determines the remaining UE-antenna positions $\mathbf{p}_{\text{u},n};\: n=2,\cdots, N$, and reconstructs $\mathbf{H}$. 
This hybrid UL/DL scheme exploits the large BS aperture for accurate first-stage estimation while reducing the UE-side task to orientation estimation only, using one UL and two DL pilot sequences. 
\subsection{UL stage: AoD Estimation at the BS}
%
In the first stage, 
the UL-active UE antenna transmits a single known UL pilot sequence of length $\tau_{\mathrm{u}}$ with transmit power $p_{\mathrm{u}}$, 
which occupies $\tau_{\mathrm{u}}$ channel uses.
Let $\mathbf{y}_\text{b}\in \mathbb{C}^M$ denote the resulting observation vector across the BS array, modeled as 
\begin{align}
\label{eq:y_b}
\mathbf{y}_\text{b}=\eta_\text{b}\mathbf{h}_{\text{b},1}(\theta_1,\theta_M)+ \mathbf{v}_\text{b},
\end{align}
where $\mathbf{h}_{\text{b},1}(\theta_1,\theta_M)\in\mathbb{C}^M$ models the channel between the BS and the first UE antenna, and its $m^\text{th}$ element is 
\begin{align}
\label{eq:h_b}
\Big(\mathbf{h}_{\text{b},1}(\theta_1,\theta_M)\Big)_{[m]}= \frac{\textit{exp}\:\Big(\frac{-\textsf{j}2\pi r_{{1,m}}(\theta_1,\theta_M)} {\lambda_\text{UL}}\Big)}{r_{1,m}(\theta_1,\theta_M)},
\end{align}
and where $\mathbf{v}_\text{b}\in\mathbb{C}^M$ denotes the effective AWGN at the BS after correlating the received signal with the single pilot sequence, where each element follows $\mathcal{CN} (0,\sigma^2_\text{b}/\tau_\text{u})$. 
Furthermore, $\eta_\text{b}\in\mathbb{C}$ denotes an unknown common complex gain that primarily absorbs the transmit-power factor and the common scaling of the LoS channel model, i.e., $\lambda_\text{UL} \sqrt{p_\text{u}}/4\pi$. It also accounts for any additional common amplitude and phase offset between $\mathbf{y}_\text{b}$ and $\mathbf{h}_{\text{b},1}$. 
To estimate the AoD pair, let $\boldsymbol{\Theta}$ denote a two-dimensional angular search space over $(\theta_1,\theta_M)$. For a given $(\theta_1,\theta_M)\in \boldsymbol{\Theta}$, we first obtain a closed-form estimate of $\eta_\text{b}$
by minimizing the squared Euclidean residual $\left\|\mathbf{y}_\text{b}-{\eta}_\text{b}\mathbf{h}_{\text{b},1}(\theta_1,\theta_M)\right\|^2$ with respect to $\eta_\text{b}$, which yields
\begin{align}
\label{eq:eta}
\hat{\eta}_\text{b}(\theta_1,\theta_M)= \frac{\mathbf{h}_{\text{b},1}^\herm(\theta_1,\theta_M)\mathbf{y}_\text{b}}{\sum_{m=1}^M r_{1,m}^{-2}(\theta_1,\theta_M)}.
\end{align}
%
Back-substituting the corresponding gain estimate from~\eqref{eq:eta} into the residual yields an optimization only over the two AoDs.  
Assuming that the BS knows its own inter-antenna spacings $l_{\text{b},m}; ~\forall m\in \setI_M$, the AoD pair is estimated by the BS as
%
\begin{equation}\label{eq:th1_thM_hat} (\hat{\theta}_1,\hat{\theta}_M)=\mathop{\text{argmin}}_{\substack{(\theta_1,\theta_M)\in \boldsymbol{\Theta}}}  \Big\|\mathbf{y}_\text{b}-\frac{\mathbf{h}_{\text{b},1}^\herm(\theta_1,\theta_M)\mathbf{y}_\text{b}\mathbf{h}_{\text{b},1}(\theta_1,\theta_M)}{\sum_{m=1}^M r_{1,m}^{-2}(\theta_1,\theta_M)}\Big\|^2.
\end{equation}
The angular grid $\boldsymbol{\Theta}$ is constructed by jointly sampling $\theta_1$ and $\theta_M$ over the feasible interval $[\theta_\text{min}, \theta_\text{max}]$. 
As discussed in \cite{Localization}, the grid efficiency can be improved by adopting a nonuniform sampling rule for at least one of the geometric reference parameters, thereby accounting for the range-dependent characteristics of the N-F wavefront. 
Specifically, in our parametrization, the candidate positions of the reference UE antenna associated with $(\theta_1,\theta_M)\in \boldsymbol{\Theta}$ should be distributed more densely at short ranges and more sparsely at larger distances. One possible construction of such a nonuniform grid, inspired by~\cite{Localization}, is provided in Appendix~I.

\subsection{DL stage: Orientation Vector Estimation at the UE}

To initiate the second estimation phase, the BS quantizes $(\hat{\theta}_1,\hat{\theta}_M)$ and conveys them in the DL. Let $Q_q(.)$ and $Q_q^{-1} (.)$ denote the $q$-bit quantization and corresponding dequantization functions, respectively. The UE uses their dequantized versions as side information, given by
\begin{align}
\label{eq:AoD_tilde}
[\tilde{\theta}_1,\tilde{\theta}_M]=\Big[Q_{q_1}^{-1} \big( Q_{q_1}(\hat{\theta}_1)\big), Q_{q_2}^{-1} \big( Q_{q_2}(\hat{\theta}_M)\big)\Big].
\end{align}
Moreover, although all BS antennas are available for DL pilot transmission, only a subset $\setM\subset\setI_M$ is selected, as in~\cite{Shima}, and the corresponding antenna-selection pattern is also provided to the UE. Similar to \cite{Shima}, $\setM$ always includes the two outermost elements, i.e., $|\setM|\geq 2$, $\setM(1)=1$, and $\setM(|\setM|)=M$.
%
%
Note that, since the reference UE-antenna position is already determined from $(\hat{\theta}_1,\hat{\theta}_M)$, the number of BS antennas required for DL pilot transmission can be reduced to as few as $|\setM|=2$, compared with the DL-pilot-only N-F LoS CSI acquisition scheme in~\cite{Shima}.
%
%
%
Let $\mathbf{y}_{\text{u},m}$ denote the observation vector associated with the $m^\text{th}$ pilot-transmitting BS antenna, $m\in \setM$, obtained at the UE after correlating the received signal with the corresponding DL pilot sequence. The $|\mathcal{M}|$ pilot-transmitting BS antennas simultaneously transmit mutually orthogonal pilot sequences of length $\tau_{\mathrm b}$, each normalized to have energy $\tau_{\mathrm b}$. 
The total BS transmit power $p_{\mathrm b}$ is equally divided among the pilot-transmitting antennas, such that each antenna transmits with power $p_{\mathrm b}/|\mathcal M|$. Hence, the DL pilot transmission occupies $\tau_{\mathrm b}$ channel uses and has a total energy of $\tau_{\mathrm b}p_{\mathrm b}$.
Accordingly,
\begin{align}
\label{eq:y_u,m}
\mathbf{y}_{\text{u},m}=\eta_\text{u}\mathbf{h}_{\text{u},m}(\boldsymbol{\psi}, {\theta}_1,{\theta}_M)+ \mathbf{v}_{\text{u},m},
\end{align}
 where $\mathbf{v}_{\text{u},m}\in\mathbb{C}^N $ is effective AWGN at the UE with each element following $\mathcal{CN} (0,\sigma^2_\text{u}/\tau_\text{b})$, 
 and $\mathbf{h}_{\text{u},m}(\boldsymbol{\psi}, {\theta}_1,{\theta}_M)\in\mathbb{C}^N$ models the channel from  $m^\text{th}$ BS antenna and the UE,  whose $n^\text{th}$ element is given by 
\begin{align}
\label{eq:h_u}
\Big(\mathbf{h}_{\text{u},m}(\boldsymbol{\psi}, {\theta}_1,{\theta}_M)\Big)_{[n]}= \frac{\textit{exp}\:\Big(\frac{-\textsf{j}2\pi r_{{n,m}}(\boldsymbol{\psi}, {\theta}_1,{\theta}_M)} {\lambda_\text{DL}}\Big)}{r_{n,m}(\boldsymbol{\psi}, {\theta}_1,{\theta}_M)}.
\end{align}
 Furthermore, $\eta_\text{u}\in \mathbb{C}$ denotes an unknown common complex gain that primarily absorbs the DL transmit-power factor and the common scaling of the LoS channel model, i.e., $\lambda_\text{DL} \sqrt{p_\text{b}}/(4\pi \sqrt{|\setM|})$. 
 It also provides common amplitude- and phase-mismatch compensation between the collected UE observations and the model-based channel response used for estimation, as detailed below.
 Stacking the observations over all $m\in \setM$ yields
\begin{align}
\label{eq:y_u}
\mathbf{y}_{\text{u}}=\eta_\text{u}\mathbf{h}_{\text{u}}(\boldsymbol{\psi}, {\theta}_1,{\theta}_M)+ \mathbf{v}_{\text{u}} \in \mathbb{C}^{N|\setM|},
\end{align}
where $\mathbf{y}_{\text{u}}=[\mathbf{y}_{\text{u},m}]_{m\in\setM}$, $\mathbf{h}_{\text{u}}=[\mathbf{h}_{\text{u},m}]_{m\in\setM}$, and $\mathbf{v}_{\text{u}}=[\mathbf{v}_{\text{u},m}]_{m\in\setM}$.
%
Since only the dequantized AoD estimates $(\tilde{\theta}_1,\tilde{\theta}_M)$ are available at the UE, it forms the channel hypothesis $\mathbf{h}_{\text{u}}(\boldsymbol{\psi}; \tilde{\theta}_1,\tilde{\theta}_M)$ for each candidate $\boldsymbol{\psi}$, which is used in the estimation process. Here, the semicolon emphasizes that $\boldsymbol{\psi}$ is the parameter to be estimated, whereas $(\tilde{\theta}_1,\tilde{\theta}_M)$ are treated as given side information. Let $\boldsymbol{\Psi}$ denote the search space of candidate UE-array orientations. For a given $\boldsymbol{\psi}\in\boldsymbol{\Psi}$, a closed-form estimate of ${\eta}_\text{u}$ is obtained by minimizing the squared Euclidean residual $\left\|\mathbf{y}_\text{u}-{\eta}_\text{u}\mathbf{h}_{\text{u}}(\boldsymbol{\psi}; \tilde{\theta}_1,\tilde{\theta}_M)\right\|^2$ with respect to ${\eta}_\text{u}$, which yields
%
\begin{align}
\label{eq:eta_u}
\hat{\eta}_\text{u}(\boldsymbol{\psi}; \tilde{\theta}_1,\tilde{\theta}_M)= \frac{\mathbf{h}_{\text{u}}^\herm(\boldsymbol{\psi}; \tilde{\theta}_1,\tilde{\theta}_M)\mathbf{y}_\text{u}}{\sum_{m\in\setM}\sum_{n=1}^N r_{n,m}^{-2}(\boldsymbol{\psi}; \tilde{\theta}_1,\tilde{\theta}_M)},
\end{align}
%
where, $r_{n,m}(\boldsymbol{\psi}; \tilde{\theta}_1,\tilde{\theta}_M)$ is obtained by evaluating \eqref{eq:r_nm} at $(\boldsymbol{\psi},\tilde{\theta}_1,\tilde{\theta}_M)$.
%
%
 Back-substituting the corresponding gain estimate into the residual leaves an optimization only over the UE orientation vector. Assuming that the UE knows the partial BS inter-antenna spacings  $l_{\text{b},m}, \forall m\in \setM$, and its own inter-antenna spacing
$l_{\text{u},n}, \forall n\in \setI_N$, the UE array orientation vector is estimated at the UE as 
\begin{equation}\label{eq:psi} \hat{\boldsymbol{\psi}}=\mathop{\text{argmin}}_{\substack{\boldsymbol{\psi}\in \boldsymbol{\Psi}}}  \bigg\|\mathbf{y}_\text{u}-\frac{\mathbf{h}_{\text{u}}^\herm(\boldsymbol{\psi}; \tilde{\theta}_1,\tilde{\theta}_M)\mathbf{y}_\text{u}\mathbf{h}_{\text{u}}(\boldsymbol{\psi}; \tilde{\theta}_1,\tilde{\theta}_M)}{\sum_{m\in\setM}\sum_{n=1}^N r_{n,m}^{-2}(\boldsymbol{\psi}; \tilde{\theta}_1,\tilde{\theta}_M)}\bigg\|^2. 
\end{equation}
%
As discussed in~\cite [Sections \MakeUppercase{\romannumeral 3}]{Shima}, the search space $\boldsymbol{\Psi}$ consists of a finite set of candidate unit vectors sampled on the unit sphere. 
Since $\boldsymbol{\psi}$ is parameterized by the two angles $(\varphi,\omega)$, i.e.,
$\boldsymbol{\psi}=\boldsymbol{\psi}(\varphi,\omega)$ as defined in~\eqref{eq:psi_vector}, the search in~\eqref{eq:psi} can equivalently be implemented as a joint search over these two angles, yielding $\hat{\boldsymbol{\psi}}=\boldsymbol{\psi}(\hat\varphi,\hat\omega)$. 
If the BS and UE arrays lie in the same plane, this representation reduces to a one-dimensional angular grid over $\varphi$, and \eqref{eq:psi} simplifies to estimating $\varphi$.
Finally, given known $l_{\text{b},m}; \forall m\in \setI_M$ and $l_{\text{u},n}; \forall n\in \setI_N$ at the BS, the BS reconstructs the channel $\hat{\mathbf{H}}$ by calculating $\hat{r}_{n,m}=r_{n,m}(\tilde{\boldsymbol{\psi}},\hat{\theta}_1,\hat{\theta}_M); \forall n\in \setI_N, m\in \setI_M$, where $\tilde{\boldsymbol{\psi}}$ denotes the received UL feedback after de-quantization. Algorithm~\ref{alg1} summarizes the above procedure.

\begin{algorithm}
  \textbf{Data:} Pilot sequences, angular spaces $\boldsymbol{\Theta}$ and $\boldsymbol{\Psi}$, the set $\setM$, $l_{\text{b},m};~ \forall m\in \setI_M$, $l_{\text{u},n};~ \forall n\in \setI_N$.    
 
 \begin{itemize}[leftmargin=12mm]
\item[\texttt{(S.0)}]
\textbf{UE}: Send the number of active antennas for UL and DL, and 
$l_{\text{u},n};~ \forall n\in\setI_N$ via an UL control channel,

\item[\texttt{(S.1)}]
\textbf{UE}: Transmit one UL pilot sequence (e.g., SRS) of length $\tau_\text{u}$ from a single antenna.  

\item[\texttt{(S.2)}]\textbf{BS}: Estimate the reference AoDs $(\theta_1, \theta_M)$ from~\eqref{eq:th1_thM_hat}.
\item[\texttt{(S.3)}]\textbf{BS}: Quantize $(\hat\theta_1, \hat\theta_M)$ and send them in the DL.
 \item[\texttt{(S.4)}]\textbf{BS}: Send $l_{\text{b},m};~ \forall m\in\setI_M$  via a DL control channel.
\item[\texttt{(S.5)}]\textbf{BS}: Transmit mutually orthogonal DL pilot sequences (e.g., CSI-RS) of length $\tau_\text{b}$ from antennas in $\setM$.
\item[\texttt{(S.6)}]
 \textbf{UE}: Fix de-quantized AoDs as $(\tilde{\theta}_1,\tilde{\theta}_M)$ via received DL signaling and estimate $\boldsymbol{\psi}$ from~\eqref{eq:psi}.
 
\item[\texttt{(S.7)}]\textbf{UE}: Quantize $\hat{\boldsymbol{\psi}}$ and feedback it in the UL.
\item[\texttt{(S.8)}] 
 \textbf{BS}: Calculate $\hat{r}_{n,m};~\forall n\in\setI_N, m\in \setI_M$ via local AoD estimates $(\hat\theta_1, \hat\theta_M)$ and de-quantized UL feedback $\tilde{\boldsymbol{\psi}}$ from~\eqref{eq:p_u,1}-\eqref{eq:r_nm}. 
 \end{itemize}
\caption{Two-stage N-F LoS MIMO channel parameter estimation under UE transmit-receive antenna asymmetry.} \label{alg1}
\end{algorithm}
\subsection {BS–UE Signaling for the Parameter Estimation}
\begin{figure}[t!]
    \centering   \includegraphics[scale=0.3]{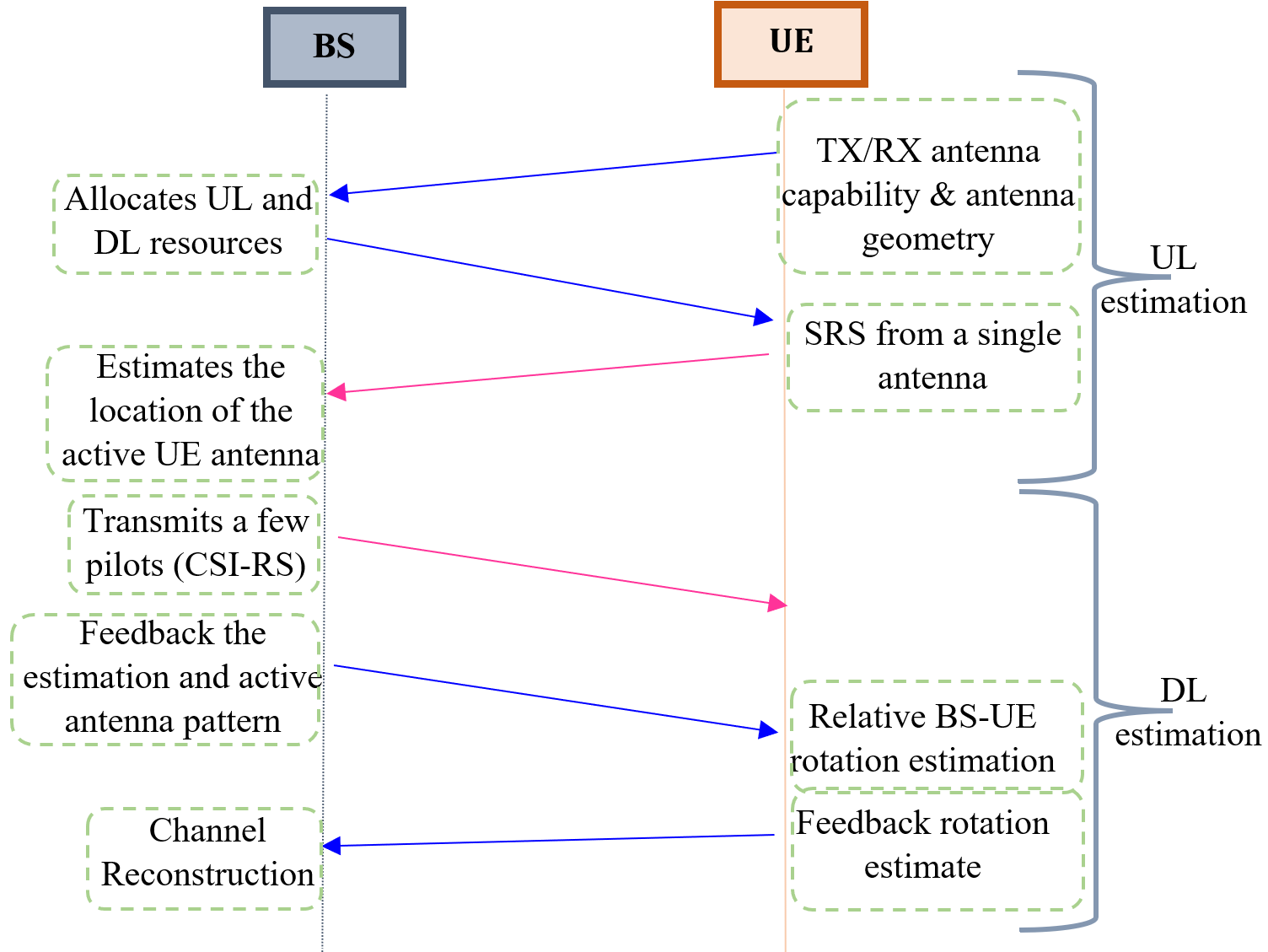} 
    \caption{{Signaling flow diagram of the proposed N-F LoS MIMO CSI acquisition.}}
    \label{fig:Signaling}
\end{figure}

As depicted in Fig.~\ref{fig:Signaling}, the channel estimation process is initiated by exchanging control information between the BS and UE, as well as pilot resource allocation. 
The UE first reports its transmit and receive antenna capability (e.g., one transmit and N receive antennas). 
It also sends the $l_{\text{u},n}, \forall n\in \setI_N$
to the BS, which is used in the channel reconstruction phase. Then, the BS configures the resources for UL and DL pilot transmissions (e.g., sounding reference signal (SRS) for UL and CSI-RS for DL), as well as the geometry-related information for the DL pilot transmission (e.g., $\setM$ and/or $l_{\text{b},m}, \forall m\in\setM$). 
%
%
The UL estimation stage begins when the UE transmits one UL pilot sequence, e.g., an SRS, from its reference antenna.
The BS collects the SRS across its antenna array, estimates $({\theta}_1,{\theta}_M)$ as in \eqref{eq:th1_thM_hat}, and sends the quantized AoD estimates~\eqref{eq:AoD_tilde} along with the pattern of its pilot-transmitting antennas (i.e., information about $\setM$). 
It then transmits CSI-RS from the selected antennas. Based on the received AoD estimates and the DL pilot observations collected across all UE antennas, the UE estimates the (relative) orientation vector $\boldsymbol{\psi}$.
Finally, the full LoS MIMO channel is reconstructed at the BS using the local estimates of AoDs, the de-quantized rotation feedback from the UE, and the known antenna spacing of both arrays.
\subsection{Computational Complexity}
Here, we analyze the computational complexity of the proposed two-stage LoS MIMO channel-parameter estimation method and compare it with the corresponding one-stage scheme in~\cite{Shima}. From~\eqref{eq:th1_thM_hat}, the dominant complexity of computing the squared-norm expression for each point of $\boldsymbol{\Theta}$ is $\mathcal{O}(M)$. Therefore, the overall complexity of the UL estimation stage, performed at the BS, is $\mathcal{O}(|\boldsymbol{\Theta}|M)$. Similarly, the complexity of the DL estimation stage in~\eqref{eq:psi}, performed at the UE, is $\mathcal{O}(|\boldsymbol{\Psi}| |\setM|N)$. Hence, the total complexity of the proposed UL/DL procedure is $\mathcal{O}\Big(|\boldsymbol{\Psi}||\setM|N+|\boldsymbol{\Theta}|M \Big)$. 
For comparison, the DL-only scheme in~\cite{Shima} performs a joint search over $(\theta_1, \theta_M,\boldsymbol{\psi})$, leading to a complexity of $\mathcal{O}(|\boldsymbol{\Theta}||\boldsymbol{\Psi}| |\setM|N)$. 
Thus, by decomposing the estimation task into BS-side AoD estimation and UE-side orientation estimation, the proposed procedure reduces the search complexity while distributing the computational effort between the BS and the UE.
\section{Estimation Performance Analysis}
\label{sec:CRLB}
%
In this section, we analyze the estimation accuracy of the proposed two-stage framework by characterizing the uncertainty of the estimated parameters. The analysis is divided into two parts. First, we derive the CRLB for the first-stage (i.e., UL phase) estimation problem, which provides a lower bound on the covariance of any locally unbiased estimator \cite{kay1993fundamentals}. Then, to account for the uncertainty of the first-stage estimates in the second stage (i.e., DL phase), we use a local first-order perturbation analysis around the true parameter values. This type of analysis is commonly used to characterize error propagation~\cite{EP_Bjorck1996, EP_Chapter9_all, EP_h_linearize, EP_Residual_approx,  EP_Boaz_Porat, EP_Sensitivity_Music_gamma;theta, EP_MIMO_radar_sensitivity_gamma;beta}. In our case, it relates the perturbation of the second-stage estimate to the estimation error inherited from the first stage and provides an approximation of its resulting covariance. 
In the ideal case where the first-stage parameters are perfectly known, the resulting expression reduces to the conventional CRLB-based covariance approximation for the second-stage parameters.
For clarity, the analysis in this section assumes perfect quantization and dequantization of the UL and DL exchanged estimation. The impact of quantization errors is incorporated separately at the end of the section.
\subsection{UL Stage: CRLB Analysis for AoDs}
Here, we analyze the accuracy of AoD estimates at the BS from the UL observation vector in~\eqref{eq:y_b}.  
The real-valued parameter vector consists of $(\theta_1, \theta_M, \Re{(\eta_\text{b})}, \Im (\eta_\text{b}))$, where $(\theta_1, \theta_M)$ are the parameters of interest and the complex gain $\eta_\text{b}$ is treated as a nuisance parameter. For a given $(\theta_1, \theta_M, \eta_\text{b})$ the UL observation follows $\mathbf{y}_\text{b} \sim \mathcal{CN} \big(\eta_\text{b}\mathbf{h}_{\text{b},1},\frac{\sigma^2_\text{b}}{\tau_\text{u}} \mathbf{I}_M\big)$. After computing the Fisher information matrix (FIM) for the joint AoD and complex-gain parameters, the nuisance gain $\eta_\text{b}$ is eliminated using the Schur complement. This yields the equivalent FIM for the parameter vector $\boldsymbol{\theta}\triangleq[\theta_1, \theta_M]^\tran$ 
as 
\begin{equation}
\label{eq:F_nub}   \mathbf{F}_{\boldsymbol{\theta}} = \frac{2\tau_\text{u} |\eta_\text{b}|^2} {\sigma^2_\text{b}}\Re\Big(\mathbf{J}_{\boldsymbol{\theta}}^\herm\mathbf{P}_\text{b}\mathbf{J}_{\boldsymbol{\theta}}\Big),
\end{equation}
where 
\begin{equation}
\label{eq:J_b}   \mathbf{J}_{\boldsymbol{\theta}}\triangleq\bigg[\frac{\partial \mathbf{h}_{\text{b},1}(\theta_1, \theta_M)}{\partial \theta_1}, \frac{\partial \mathbf{h}_{\text{b},1}(\theta_1, \theta_M)}{\partial \theta_M}\bigg]
\end{equation}
\begin{equation}
\label{eq:P_b} 
\mathbf{P}_\text{b} =\mathbf{I}_M-\frac{\mathbf{h}_{\text{b},1}(\theta_1, \theta_M)\mathbf{h}_{\text{b},1}^\herm(\theta_1, \theta_M)}{\|\mathbf{h}_{\text{b},1}(\theta_1, \theta_M)\|^2}.
\end{equation}
Proof of~\eqref{eq:F_nub} is provided in the Appendix \ref{app:A1}.
Partial derivatives appearing in \eqref{eq:J_b} are expressed as
\begin{equation}
\label{eq:doh_hb_doh_thet}
\frac{\partial \mathbf{h}_{\text{b},1}(\theta_1, \theta_M)}{\partial \theta_i}=\mathbf{h}_{\text{b},1}(\theta_1, \theta_M)\odot\mathbf{w}_{\theta_i}(\theta_1, \theta_M) ,\quad i=1,M,
\end{equation}
where $\odot$ is the element-wise product and the $m^\text{th}$  element of $\mathbf{w}_{\theta_i}\in \mathbb{C}^{M\times1}$ is given as
\begin{equation}
\label{eq:g_b}
\big(\mathbf{w}_{\theta_i}\big)_{[m]}=\frac{-\partial r_{1,m}}{\partial \theta_i}\big(\frac{2\pi\textsf{j}}{\lambda_{\text{UL}}}+\frac{1}{r_{1,m}}\big), m\in \setI_M;  i=1,M,
\end{equation}
where the arguments $(\theta_1, \theta_M)$ have been omitted from $\mathbf{w}_{\theta_i}$ and $r_{1,m}$ for notational simplicity.
The closed-form expression of $r_{1,m}$ is obtained from \eqref{eq:p_u,1}-\eqref{eq:r_nm} as
\begin{equation}
\label{eq:r_1,m}
r_{1,m}=\sqrt{r_{1,1}^2+l^2_{\text{b},m}-2r_{1,1}l_{\text{b},m}\sin\theta_1}. 
\end{equation}
Thus, derivative terms in~\eqref{eq:g_b} are given by
\begin{equation}
\label{eq:doh_r_doh_thet1}
\frac{\partial r_{1,m}}{\partial \theta_1}=\frac{r_{1,1}}{r_{1,m}}\Big(\cot(\theta_M-\theta_1)(r_{1,1}-l_{\text{b},m} \sin\theta_1)+\cos\theta_1\Big),
\end{equation}
\begin{equation}
\label{eq:doh_r_doh_thetM}
\frac{\partial r_{1,m}}{\partial \theta_M}=\frac{l_{\text{b},M}\cos\theta_1}{r_{1,m}\sin^2(\theta_M-\theta_1)}\Big(l_{\text{b},m}\sin\theta_1-r_{1,1}\Big).
\end{equation}
For any locally unbiased AoD estimator, the error covariance matrix satisfies $\mathbb{E} \{\boldsymbol{e}_{\boldsymbol{\theta}}\boldsymbol{e}_{\boldsymbol{\theta}}^\tran\}\succeq\mathbf{F}_{\boldsymbol{\theta}}^{-1}$, where $\boldsymbol{e}_{\boldsymbol{\theta}}\triangleq\hat{\boldsymbol{\theta}}-\boldsymbol{\theta}$. Accordingly, the MSE of each AoD is lower bounded as $(\mathbf{F}_{\boldsymbol{\theta}}^{-1})_{[1,1]}\leq\Exp\Big\{\big| \hat{\theta}_{1}-\theta_{1}\big|^2\Big\}$ and $(\mathbf{F}_{\boldsymbol{\theta}}^{-1})_{[2,2]}\leq\Exp\Big\{\big| \hat{\theta}_{M}-\theta_{M}\big|^2\Big\}$. 
\subsection{DL Stage: Error Propagation Analysis for UE Orientation}
\label{DL_error}
For the DL-stage error analysis, we parameterize the UE orientation vector $\boldsymbol{\psi}$ through its two independent degrees of freedom, $\boldsymbol{\nu}\triangleq[\varphi,\omega]^\tran$, such that $\boldsymbol{\psi}=\boldsymbol{\psi}(\varphi,\omega)$.The analysis is then carried out with respect to these two orientation parameters.
Under this parameterization, $\mathbf{h}_\text{u}(\boldsymbol{\psi};\theta_1,\theta_M)$ in \eqref{eq:y_u} is equivalently written as $\mathbf{h}_\text{u}(\varphi, \omega;\theta_1, \theta_M)=\mathbf{h}_\text{u}(\boldsymbol{\nu};\boldsymbol{\theta})$. 

\subsubsection{Perfect AoDs}
\label{perfect_AoD}
 In the ideal case, the UL parameters $(\theta_1,\theta_M)$ are perfectly known at the UE. Accordingly, $(\tilde{\theta}_1, \tilde{\theta}_M)$ is replaced by $(\theta_1,\theta_M)$ in \eqref{eq:psi}. 
For given $\boldsymbol{\theta}$, $\boldsymbol{\nu}$, and $\eta_\text{u}$, the DL observation vector received at the UE follows $\mathbf{y}_\text{u} \sim \mathcal{CN} \big(\eta_\text{u}\mathbf{h}_\text{u},\frac{\sigma^2_\text{u} }{\tau_\text{b}} \mathbf{I}_{N|\setM|}\big)$. Treating $\eta_\text{u}$ as a nuisance parameter, the equivalent FIM for the UE orientation parameters is derived following the same procedure as the UL-stage AoD FIM in Appendix~\ref{app:A1}, yielding
\begin{equation}
\label{eq:F_rot}   \mathbf{F}_{\boldsymbol{\nu}}= \frac{2 \tau_\text{b}  |\eta_\text{u}|^2} {\sigma^2_\text{u}}\Re\Big(\mathbf{D}_{\boldsymbol{\nu}}^\herm\mathbf{P}_\text{u}\mathbf{D}_{\boldsymbol{\nu}}\Big),
\end{equation}
where 
\begin{equation}
\label{eq:J_u}   
\mathbf{D}_{\boldsymbol{\nu}}=\bigg[\frac{\partial \mathbf{h}_{\text{u}}(\boldsymbol{\nu};\boldsymbol{\theta})}{\partial \boldsymbol{\nu}_{[1]}}, \frac{\partial \mathbf{h}_{\text{u}}(\boldsymbol{\nu};\boldsymbol{\theta})}{\partial \boldsymbol{\nu}_{[2]}}\bigg], 
\end{equation}
\begin{equation}
\label{eq:P_u} 
\mathbf{P}_\text{u} =\mathbf{I}_{N|\setM|}-\frac{\mathbf{h}_{\text{u}}(\boldsymbol{\nu};\boldsymbol{\theta})\mathbf{h}_{\text{u}}^\herm(\boldsymbol{\nu};\boldsymbol{\theta})}{\|\mathbf{h}_{\text{u}}(\boldsymbol{\nu};\boldsymbol{\theta})\|^2},
\end{equation}
The partial derivative terms in \eqref{eq:J_u} are given as 
\begin{equation}
\label{eq:doh_hu_doh_rot}
\frac{\partial \mathbf{h}_{\text{u}}(\boldsymbol{\nu};\boldsymbol{\theta})}{\partial \boldsymbol{\nu}_{[i]}}=\mathbf{h}_{\text{u}}(\boldsymbol{\nu};\boldsymbol{\theta})\odot\mathbf{w}_{\boldsymbol{\nu}_{[i]}}(\boldsymbol{\nu};\boldsymbol{\theta}) ,\quad i=1,2,
\end{equation}
where $\mathbf{w}_{\boldsymbol{\nu}_{[i]}}=[\mathbf{w}_{m,\boldsymbol{\nu}_{[i]}}]_{m\in\setM}$, in which the $n^\text{th}$ element of $\mathbf{w}_{m,\boldsymbol{\nu}_{[i]}}\in \mathbb{C}^N$ is given by
\begin{equation}
\label{eq:g_u}
\big(\mathbf{w}_{m,\boldsymbol{\nu}_{[i]}})_{[n]}=\frac{-\partial r_{n,m}}{\partial \boldsymbol{\nu}_{[i]}}\big(\frac{2\pi\textsf{j}}{\lambda_\text{DL}}+\frac{1}{r_{n,m}}\big), n\in \setI_N;  i=1,2,
\end{equation}
where $r_{n,m}$ is given in closed-form by~\eqref{r_n,m_given_r1m} (on the top of the next page).
\begin{figure*}[t!]
\begin{align}\label{r_n,m_given_r1m}
r_{n,m}=\sqrt{r_{1,m}^2+l^2_{\text{u},n}-2l_{\text{u},n}\sin\omega(r_{1,1}\sin(\varphi+\theta_1)-l_{\text{b},m}\cos\varphi)},
\end{align}
\hrulefill
\end{figure*}
Let $\hat{\boldsymbol{\nu}}^{(0)}$ denote the estimated orientation parameter vector when the AoDs are perfectly known. For any locally unbiased estimator, the error covariance matrix satisfies $\mathbb{E} \{\boldsymbol{e}_{\boldsymbol{\nu}}^{(0)}(\boldsymbol{e}_{\boldsymbol{\nu}}^{(0)})^\tran\}\succeq\mathbf{F}_{\boldsymbol{\nu}}^{-1}$, where $\boldsymbol{e}_{\boldsymbol{\nu}}^{(0)}\triangleq\hat{\boldsymbol{\nu}}^{(0)}-\boldsymbol{\nu}$. Hence, with perfect knowledge of the AoDs, the MSE of each orientation angle (i.e., azimuth and elevation) is lower bounded as $(\mathbf{F}_{\boldsymbol{\nu}}^{-1})_{[i, i]}\leq\Exp\Big\{\big| \hat{\boldsymbol{\nu}}^{(0)}_{[i]}-\boldsymbol{\nu}_{[i]}\big|^2\Big\};~i=1,2$.
\subsubsection{Erroneous AoDs}
\label{Erroneous AoDs}
We now consider imperfect AoD estimates inherited from the UL stage and analyze how their errors propagate to the DL orientation estimate.
To simplify the derivation, we first assume that the complex gain parameters $\eta_\text{u}$ and $\eta_\text{b}$ are perfectly known. Following the discussion in the Appendix~\ref{app:A1}, the impact of the unknown gain parameters is incorporated at the end of the derivation through the derivative matrices modified using the previously introduced projection matrices $\mathbf{P}_\text{u}$ and $\mathbf{P}_\text{b}$. Thus, we focus on the estimation problem
\begin{align}
\label{eq:nu_hat}
\hat{\boldsymbol{\nu}}&= \mathop{\text{argmin}}_{\substack{\boldsymbol{\nu}\in \mathbb{R}^2}} \|\mathbf{y}_{\text{u}}-\eta_\text{u}\mathbf{h}_\text{u}(\boldsymbol{\nu};\hat{\boldsymbol{\theta}})\|^2,
\end{align}
which is equivalent to \eqref{eq:psi} when $\eta_\text{u}$ assumed to be known and the orientation vector $\boldsymbol{\psi}$ is parametrized through the parameter vector $\boldsymbol{\nu}$. 
To derive the error propagation, we use a first-order local perturbation analysis around the true parameter values, a standard technique for characterizing estimation-error propagation~\cite{EP_Bjorck1996,EP_Chapter9_all,EP_h_linearize,EP_Residual_approx, EP_Boaz_Porat,EP_Sensitivity_Music_gamma;theta, EP_MIMO_radar_sensitivity_gamma;beta}. For sufficiently small estimation errors, the nonlinear channel model $\mathbf{h}_\text{u}(\hat{\boldsymbol{\nu}};\hat{\boldsymbol{\theta}})$ can be locally approximated by its first-order Taylor expansion, which transforms \eqref{eq:nu_hat} into a linear least-squares problem for the estimation error of $\boldsymbol{\nu}$. 
To obtain this approximation, we define $\boldsymbol{e}_{\boldsymbol{\nu}}=\hat{\boldsymbol{\nu}}-\boldsymbol{\nu}$ as the estimation error of the orientation-parameter vector and recall the previously defined AoD estimation error $\boldsymbol{e}_{\boldsymbol{\theta}}=\hat{\boldsymbol{\theta}}-\boldsymbol{\theta}$.
The first-order Taylor approximation of $\mathbf{h}_\text{u}(\hat{\boldsymbol{\nu}};\hat{\boldsymbol{\theta}})$ around the true parameter values is 
\begin{align}
\label{eq:h_u_apprx}
\mathbf{h}_{\text{u}}(\hat{\boldsymbol{\nu}};\hat{\boldsymbol{\theta}})\approx\mathbf{h}_{\text{u}}({\boldsymbol{\nu}};{\boldsymbol{\theta}})+\mathbf{D}_{\boldsymbol{\nu}}({\boldsymbol{\nu}};{\boldsymbol{\theta}})\boldsymbol{e}_{\boldsymbol{\nu}}+\mathbf{D}_{\boldsymbol{\theta}}({\boldsymbol{\nu}};{\boldsymbol{\theta}})\boldsymbol{e}_{\boldsymbol{\theta}}, 
\end{align}
where $\mathbf{D}_{\boldsymbol{\nu}}$ defined in \eqref{eq:J_u}, is compactly expressed as
\begin{align}
\label{eq:D_nu}
\mathbf{D}_{\boldsymbol{\nu}}(\boldsymbol{\nu};{\boldsymbol{\theta}})=\frac{\partial\mathbf{h}_{\text{u}}(\boldsymbol{\nu};{\boldsymbol{\theta}})}{\partial\boldsymbol{\nu}^\tran}\in \mathbb{C}^{N|\setM|\times 2}, 
\end{align}
and
\begin{align}
\label{eq:D_thet}
\mathbf{D}_{\boldsymbol{\theta}}(\boldsymbol{\nu};{\boldsymbol{\theta}})=\frac{\partial\mathbf{h}_{\text{u}}(\boldsymbol{\nu};{\boldsymbol{\theta}})}{\partial\boldsymbol{\theta}^\tran}\in \mathbb{C}^{N|\setM|\times 2}. 
\end{align}
Now, we define the residual vector of the DL-stage estimation problem~\eqref{eq:nu_hat}, evaluated at $\hat{\boldsymbol{\nu}}$ as
\begin{align}
\label{eq:y_res}
\mathbf{y}_{\text{res}}(\hat{\boldsymbol{\nu}};\hat{\boldsymbol{\theta}})\triangleq\mathbf{y}_\text{u}-\eta_\text{u}\mathbf{h}_\text{u}(\hat{\boldsymbol{\nu}};\hat{\boldsymbol{\theta}}),
\end{align}
Substituting \eqref{eq:h_u_apprx} in \eqref{eq:y_res} and using the DL observation model in~\eqref{eq:y_u} to write $\mathbf{y}_{\text{u}}-\eta_\text{u}\mathbf{h}_{\text{u}}({\boldsymbol{\nu}};{\boldsymbol{\theta}})=\mathbf{v}_{\text{u}}$ yields
\begin{equation}
\begin{aligned}
\label{eq:res_apprx}
\mathbf{y}_{\text{res}}(\hat{\boldsymbol{\nu}};\hat{\boldsymbol{\theta}})&\approx\mathbf{v}_{\text{u}}-\eta_\text{u}\mathbf{D}_{\boldsymbol{\nu}}({\boldsymbol{\nu}};{\boldsymbol{\theta}})\boldsymbol{e}_{\boldsymbol{\nu}}-\eta_\text{u}\mathbf{D}_{\boldsymbol{\theta}}({\boldsymbol{\nu}};{\boldsymbol{\theta}})\boldsymbol{e}_{\boldsymbol{\theta}}\\&=\mathbf{b}-\eta_\text{u}\mathbf{D}_{\boldsymbol{\nu}}({\boldsymbol{\nu}};{\boldsymbol{\theta}})\boldsymbol{e}_{\boldsymbol{\nu}},
\end{aligned}
\end{equation}
where $\mathbf{b}\triangleq \mathbf{v}_{\text{u}}-\eta_\text{u}\mathbf{D}_{\boldsymbol{\theta}}({\boldsymbol{\nu}};{\boldsymbol{\theta}})\boldsymbol{e}_{\boldsymbol{\theta}}$ 
%
From\eqref{eq:res_apprx}, minimizing the squared Euclidean norm of the linearized residual with respect to the real-valued vector $\boldsymbol{e}_{\boldsymbol{\nu}}$ yields a closed-form approximation for $\boldsymbol{e}_{\boldsymbol{\nu}}$, given by
%
\begin{equation}
\begin{aligned}
\label{eq:e_nu_LS}
\boldsymbol{e}_{\boldsymbol{\nu}}&\approx \mathop{\text{argmin}}_{\substack{\boldsymbol{e}_{\boldsymbol{\nu}}\in \mathbb{R}^2}} \|\mathbf{b}-\eta_\text{u}\mathbf{D}_{\boldsymbol{\nu}}\boldsymbol{e}_{\boldsymbol{\nu}}\|^2\\&=\Big(|\eta_\text{u}|^2\Re\big(\mathbf{D}_{\boldsymbol{\nu}}^\herm\mathbf{D}_{\boldsymbol{\nu}}\big)\Big)^{-1}\Re\big(\eta_\text{u}^*\mathbf{D}_{\boldsymbol{\nu}}^\herm\mathbf{b}\big), 
\end{aligned}
 \end{equation}
where the arguments $({\boldsymbol{\nu}};{\boldsymbol{\theta}})$ of $\mathbf{D}_{\boldsymbol{\nu}}$ are omitted for brevity.
To facilitate a FIM-based representation, we multiply both the matrix and vector terms on the right-hand side of \eqref{eq:e_nu_LS} by the common factor $2\sigma^{-2}$, where $\sigma^{-2}\triangleq \sigma_\text{u}^{-2}\tau_\text{b}$, which leaves the solution unchanged.  Accordingly 
\begin{equation}
\begin{aligned}
\label{eq:A'A}
\tilde{\mathbf{F}}_{\boldsymbol{\nu}}\triangleq2\sigma^{-2}|\eta_\text{u}|^2\Re\Big(\mathbf{D}^\herm_{\boldsymbol{\nu}}(\boldsymbol{\nu};\boldsymbol{\theta})\mathbf{D}_{\boldsymbol{\nu}}(\boldsymbol{\nu};\boldsymbol{\theta})\Big),
\end{aligned}
\end{equation}
%
%
where, similarly to $\mathbf{F}_{\boldsymbol{\nu}}$ in \eqref{eq:F_nu_thet}, $\tilde{\mathbf{F}}_{\boldsymbol{\nu}}$ represents the FIM of $\boldsymbol{\nu}$, under the assumption that $\boldsymbol{\theta}$ and $\eta_\text{u}$ are perfectly known.
From a local sensitivity perspective, 
$\tilde{\mathbf{F}}_{\boldsymbol{\nu}}$ may also be viewed as a self-sensitivity matrix that quantifies how strongly the observation model $\mathbf{h}_\text{u}$ varies with perturbations in $\boldsymbol{\nu}$. 
Moreover,
\begin{equation}
\label{eq:A'b}
\begin{aligned}
2\sigma^{-2}\Re(\eta_\text{u}^*\mathbf{D}_{\boldsymbol{\nu}}^{\herm}\mathbf{b})&=2\sigma^{-2}\Re\Big(\eta_\text{u}^*\mathbf{D}^\herm_{\boldsymbol{\nu}}(\boldsymbol{\nu};\boldsymbol{\theta})\mathbf{v}_\text{u}\Big)\\&-2\sigma^{-2}|\eta_\text{u}|^2\Re\Big(\mathbf{D}^\herm_{\boldsymbol{\nu}}(\boldsymbol{\nu};\boldsymbol{\theta})\mathbf{D}_{\boldsymbol{\theta}}(\boldsymbol{\nu};\boldsymbol{\theta})\Big)\boldsymbol{e}_{\boldsymbol{\theta}}\\&\triangleq\tilde{\mathbf{v}}_{\text{u}}-\tilde{\mathbf{S}}_{\boldsymbol{\nu}\boldsymbol{\theta}}\boldsymbol{e}_{\boldsymbol{\theta}},
\end{aligned}
\end{equation}
where $\tilde{\mathbf{v}}_\text{u}$ is the effective noise term, and $\tilde{\mathbf{S}}_{\boldsymbol{\nu}\boldsymbol{\theta}}$ can be interpreted as a cross-sensitivity matrix that quantifies the local coupling between $\boldsymbol{\nu}$ and $\boldsymbol{\theta}$. Substituting \eqref{eq:A'A} and \eqref{eq:A'b} in \eqref{eq:e_nu_LS}, we obtain
\begin{align}
\label{eq:e_final}
\boldsymbol{e}_{\boldsymbol{\nu}}\approx\tilde{\mathbf{F}}_{\boldsymbol{\nu}}^{-1}\big(\tilde{\mathbf{v}}_{\text{u}}-\tilde{\mathbf{S}}_{\boldsymbol{\nu}\boldsymbol{\theta}}\boldsymbol{e}_{\boldsymbol{\theta}}\big).
\end{align}
 Assuming 
 that the first-stage estimation error $\boldsymbol{e}_{\boldsymbol{\theta}}$ is independent of the DL noise $\mathbf{v}_\text{u}$, the covariance matrix of $\boldsymbol{e}_{\boldsymbol{\nu}}$, under the assumption $\eta_\text{u}$ and $\eta_\text{b}$ are known, can be approximated as
 \begin{align}\label{eq:Cov_e}
\mathbb{E}\{\boldsymbol{e}_{\boldsymbol{\nu}}\boldsymbol{e}_{\boldsymbol{\nu}}^\tran\}\approx\tilde{\mathbf{F}}_{\boldsymbol{\nu}}^{-1}\Big(\mathbb{E}\{\tilde{\mathbf{v}}_{\text{u}}\tilde{\mathbf{v}}_{\text{u}}^\tran\}+\tilde{\mathbf{S}}_{\boldsymbol{\nu}\boldsymbol{\theta}}\mathbb{E}\{\boldsymbol{e}_{\boldsymbol{\theta}}\boldsymbol{e}_{\boldsymbol{\theta}}^\tran\}\tilde{\mathbf{S}}^\tran_{\boldsymbol{\nu}\boldsymbol{\theta}}\Big)\tilde{\mathbf{F}}_{\boldsymbol{\nu}}^{-1},
\end{align}
where
$\mathbb{E}\{\boldsymbol{e}_{\boldsymbol{\theta}}\boldsymbol{e}_{\boldsymbol{\theta}}^\tran\}$ is the covariance matrix of the AoD estimation error under the known $\eta_\text{b}$ assumption, and is lower-bounded by the inverse of the corresponding FIM, i.e., $\mathbb{E}\{\boldsymbol{e}_{\boldsymbol{\theta}}\boldsymbol{e}_{\boldsymbol{\theta}}^\tran\}\succeq\tilde{\mathbf{F}}_{\boldsymbol{\theta}}^{-1}$, where
 \begin{align}\label{eq:F_tilde_theta}
\tilde{\mathbf{F}}_{\boldsymbol{\theta}}=2\sigma_\text{b}^{-2}\tau_\text{u}|\eta_\text{b}|^2\Re\Big(\mathbf{J}^\herm_{\boldsymbol{\theta}}(\boldsymbol{\theta})\mathbf{J}_{\boldsymbol{\theta}}(\boldsymbol{\theta})\Big),
\end{align}
and $\mathbf{J}_{\boldsymbol{\theta}}$ is defined in \eqref{eq:J_b}. In the local regime where the first-stage estimator operates near the true solution and approaches the CRLB, we therefore approximate $\mathbb{E}\{\boldsymbol{e}_{\boldsymbol{\theta}}\boldsymbol{e}_{\boldsymbol{\theta}}^\tran\}$ in \eqref{eq:Cov_e} by $\tilde{\mathbf{F}}_{\boldsymbol{\theta}}^{-1}$.
Moreover, the covariance matrix of $\tilde{\mathbf{v}}_\text{u}$ is obtained as 
 \begin{equation}
 \begin{aligned}\label{eq:Cov_v_tild}
\mathbb{E}\{\tilde{\mathbf{v}}_\text{u}\tilde{\mathbf{v}}_\text{u}^\tran\}&=\frac{4}{\sigma^4}\mathbb{E}\Big\{\Re\Big(\eta_\text{u}^*\mathbf{D}^\herm_{\boldsymbol{\nu}}\mathbf{v}_\text{u}\Big)\Re\Big(\eta_\text{u}^*\mathbf{D}^\herm_{\boldsymbol{\nu}}\mathbf{v}_\text{u}\Big)^\tran\Big\}\\&=\frac{2}{\sigma^4}\Re\Big(\mathbb{E}\Big\{\eta_\text{u}^*\mathbf{D}^\herm_{\boldsymbol{\nu}}\mathbf{v}_\text{u}\mathbf{v}_\text{u}^\herm\mathbf{D}_{\boldsymbol{\nu}}\eta_\text{u}\Big\}\Big)\\& =\frac{2}{\sigma^4}|\eta_\text{u}|^2\Re\Big(\mathbf{D}^\herm_{\boldsymbol{\nu}}\mathbb{E}\Big\{\mathbf{v}_\text{u}\mathbf{v}_\text{u}^\herm \Big\}\mathbf{D}_{\boldsymbol{\nu}}\Big)\\&=\frac{2}{\sigma^2}|\eta_\text{u}|^2\Re\Big(\mathbf{D}^\herm_{\boldsymbol{\nu}}\mathbf{D}_{\boldsymbol{\nu}}\Big)=\tilde{\mathbf{F}}_{\boldsymbol{\nu}},
\end{aligned}
\end{equation}
where $\mathbf{D}_{\boldsymbol{\nu}}=\mathbf{D}_{\boldsymbol{\nu}}(\boldsymbol{\nu};\boldsymbol{\theta})$, and the second equality follows the fact that for any proper\footnote{According to~\cite{Proper_vec}, a complex random vector $\mathbf{x}$ is called proper if we have 
$
\mathbb{E}\left\{
(\mathbf{x}-\mathbb{E}\{\mathbf{x}\})
(\mathbf{x}-\mathbb{E}\{\mathbf{x}\})^\tran
\right\}
=\mathbf{0}
$.
For a zero-mean vector, this reduces to $\mathbb{E}\{\mathbf{x}\mathbf{x}^\tran\}=\mathbf{0}$. }
zero-mean complex random vector $\mathbf{x}$ we have $\mathbb{E}\{\Re(\mathbf{x})\Re(\mathbf{x})^\tran\}=0.5~\Re(\mathbb{E}\{\mathbf{x}\mathbf{x}^\herm\})$. 
Substituting~\eqref{eq:Cov_v_tild} in \eqref{eq:Cov_e} and and using $\tilde{\mathbf{F}}_{\boldsymbol{\theta}}^{-1}$ to approximate$\mathbb{E}\{\boldsymbol{e}_{\boldsymbol{\theta}}\boldsymbol{e}_{\boldsymbol{\theta}}^\tran\}$ gives
\begin{align}\label{eq:Cov_e_Final}
\mathbb{E}\{\boldsymbol{e}_{\boldsymbol{\nu}}\boldsymbol{e}_{\boldsymbol{\nu}}^\tran\}\approx\tilde{\mathbf{F}}_{\boldsymbol{\nu}}^{-1}+\tilde{\mathbf{F}}_{\boldsymbol{\nu}}^{-1}\tilde{\mathbf{S}}_{\boldsymbol{\nu}\boldsymbol{\theta}}\tilde{\mathbf{F}}_{\boldsymbol{\theta}}^{-1}\tilde{\mathbf{S}}^\tran_{\boldsymbol{\nu}\boldsymbol{\theta}}\tilde{\mathbf{F}}_{\boldsymbol{\nu}}^{-1}.
\end{align}
The first term in \eqref{eq:Cov_e_Final} represents the intrinsic DL-stage estimation uncertainty due to the DL observation noise, whereas the second term quantifies additional estimation uncertainty in the UE-array orientation angles due to the propagation of first-stage AoD estimation errors. If $\boldsymbol{\theta}$ is perfectly known at the UE, the uncertainty of $\boldsymbol{\theta}$ is zero and the second term vanishes, recovering the conventional CRLB-based covariance approximation for the UE-array orientation angles.
To account for the unknown gains $\eta_\text{u}$ and $\eta_\text{b}$, We project the derivative matrices onto the orthogonal complements of the corresponding channel vectors using the projection matrices $\mathbf{P}_\text{b}$ and $\mathbf{P}_\text{u}$, defined in \eqref{eq:P_b} and \eqref{eq:P_u}, respectively.
%
%
%
Specifically, 
$\mathbf{D}_{\boldsymbol{\nu}}(\boldsymbol{\nu};\boldsymbol{\theta})$ 
%
and $\mathbf{D}_{\boldsymbol{\theta}}(\boldsymbol{\nu};\boldsymbol{\theta})$ are replaced by $\mathbf{P}_\text{u}\mathbf{D}_{\boldsymbol{\nu}}(\boldsymbol{\nu};\boldsymbol{\theta})$ and $\mathbf{P}_\text{u}\mathbf{D}_{\boldsymbol{\theta}}(\boldsymbol{\nu};\boldsymbol{\theta})$, respectively, while  
$\mathbf{J}_{\boldsymbol{\theta}}(\boldsymbol{\theta})$ is replaced by $\mathbf{P}_\text{b}\mathbf{J}_{\boldsymbol{\theta}}(\boldsymbol{\theta})$.
Consequently, $\tilde{\mathbf{F}}_{\boldsymbol{\nu}}$, $\tilde{\mathbf{S}}_{\boldsymbol{\nu}\boldsymbol{\theta}}$, and $\tilde{\mathbf{F}}_{\boldsymbol{\theta}}$ become ${\mathbf{F}}_{\boldsymbol{\nu}}$, ${\mathbf{S}}_{\boldsymbol{\nu}\boldsymbol{\theta}}$, and ${\mathbf{F}}_{\boldsymbol{\theta}}$, respectively, 
 and \eqref{eq:Cov_e_Final} becomes
\begin{align}\label{eq:Cov_e_Final_theta_hat}
\mathbb{E}\{\boldsymbol{e}_{\boldsymbol{\nu}}\boldsymbol{e}_{\boldsymbol{\nu}}^\tran\}\approx{\mathbf{F}}_{\boldsymbol{\nu}}^{-1}+{\mathbf{F}}_{\boldsymbol{\nu}}^{-1}{\mathbf{S}}_{\boldsymbol{\nu}\boldsymbol{\theta}}{\mathbf{F}}_{\boldsymbol{\theta}}^{-1}{\mathbf{S}}^\tran_{\boldsymbol{\nu}\boldsymbol{\theta}}{\mathbf{F}}_{\boldsymbol{\nu}}^{-1}.
\end{align}

Finally, we account for the quantization of the AoD estimates conveyed from the BS to the UE and the orientation parameters fed back from the UE to the BS.
Let $q_1$ and $q_2$ denote the quantization bits of the elements of $\boldsymbol{\theta}$, and $q_3$ and $q_4$ those of the elements of $\boldsymbol{\nu}$. Assuming uniform quantization (and ideal dequantization), the corresponding quantization-error variance is given by $\sigma^2_{q_i}=
\frac{(a_{i,\text{min}}-a_{i,\text{max}})^2}{12~2^{2q_i}};~ i\in\setI_4$, where $a_{i,\text{min}}$ and $a_{i,\text{max}}$ are the minimum and maximum values of the $i^\text{th}$ parameter. Thus, the quantization-error covariance matrices associated with $\boldsymbol{\theta}$ and $\boldsymbol{\nu}$ are defined as $\mathbf{Q}_{\boldsymbol{\theta}}=\diag(\sigma^2_{q_1},\sigma^2_{q_2})$ and $\mathbf{Q}_{\boldsymbol{\nu}}=\diag(\sigma^2_{q_3},\sigma^2_{q_4})$, respectively. 
The AoD quantization error contributes to the input uncertainty of the second-stage estimator and is therefore incorporated through ${\mathbf{F}}_{\boldsymbol{\theta}}^{-1}+\mathbf{Q}_{\boldsymbol{\theta}}$.  
In contrast, quantization of the orientation parameters occurs during their feedback to the BS and contributes an additional covariance term $\mathbf{Q}_{\boldsymbol{\nu}}$ to the orientation-parameter uncertainty at the BS. Hence, the covariance of the orientation parameters available at the BS can be approximated as 
\begin{align}
\mathbb{E}\{\boldsymbol{e}_{\boldsymbol{\nu}}\boldsymbol{e}_{\boldsymbol{\nu}}^\tran\}\approx
{\mathbf{F}}_{\boldsymbol{\nu}}^{-1}
+{\mathbf{F}}_{\boldsymbol{\nu}}^{-1}{\mathbf{S}}_{\boldsymbol{\nu}\boldsymbol{\theta}}
\big({\mathbf{F}}_{\boldsymbol{\theta}}^{-1}+\mathbf{Q}_{\boldsymbol{\theta}}\big)
{\mathbf{S}}^\tran_{\boldsymbol{\nu}\boldsymbol{\theta}}{\mathbf{F}}_{\boldsymbol{\nu}}^{-1}
+\mathbf{Q}_{\boldsymbol{\nu}}.
\end{align}
\section{Numerical Results and Discussion}
\label{sec:num}
%
In this section, we assess the performance of the proposed two-stage N-F LoS MIMO channel-parameter estimation scheme under asymmetric UE antenna capabilities. 
The simulation setup is motivated by the indoor short-range LoS scenarios considered in~\cite{Shima} and and adopts the LoS propagation model in~\cite {Indoor_mmw}. In particular, the considered geometry is representative of indoor applications such as large meeting rooms or auditoriums, where the BS may be mounted on a side wall or ceiling and the UEs are located within the room. 
We assume a scenario where the number of BS and UE antennas is $M=64$ and $N=8$, respectively. The carrier wavelengths are set to $\lambda_\text{UL}=\lambda_\text{DL}=\lambda=0.03\,\mathrm{m}$ corresponding to $10\,\mathrm{GHz}$ frequency. The length of the BS array is fixed to $2\, \mathrm{m}$, while for the UE it is $10.5\, \mathrm{cm}$.
In the simulations, without loss of generality, we consider the coplanar case, i.e., $\omega=\pi/2$, which also allows the reuse of the BS pilot-antenna patterns reported in~\cite{Shima} for the DL-only baseline.
In this scenario, estimating $\boldsymbol{\psi}$ reduces to estimating $\varphi$, and the estimation parameters become $(\theta_1,\theta_M,\varphi)$.

We generate $1600$ random UE locations with $r_{1,1}\in [3.5\mathrm{m}, 9.5\mathrm{m}]$, $\theta_1\in[-60^\circ, 60^\circ]$
\footnote{The first BS and UE antennas are used as the reference elements
for $(\theta_1,r_{1,1})$, and correspond to the elements with the smallest
$y$-coordinates in the $xy$-plane. Since these reference elements are
located at the array ends rather than their centers, a spatially symmetric
UE region does not necessarily result in a symmetric
$(\theta_1,r_{1,1})$ parameter region.}
, $\omega=90^\circ$ and $\varphi\in[0^\circ, 180^\circ]$.
%
For DL pilot transmission of the proposed scheme, the two active BS antennas transmit mutually orthogonal pilot sequences constructed from a DFT matrix, with pilot length $\tau_\text{b}=4$.
For the UL pilot transmission, we use a single pilot sequence of length $\tau_\text{u}=1$. 
The noise variances are set to $\sigma^2_\text{b}=\sigma^2_\text{u}=-85\,\text{dBm}$,  while the UL and DL transmit powers $p_\text{u}$ and $p_\text{b}$ are specified for each experiment. 
For the search-based estimation, the AoD search space $\boldsymbol{\Theta}$ is constructed according to the procedure described in the Appendix~\ref{app:first}. In the simulations, $\theta_1$ is uniformly sampled over $[-\pi/2, \pi/2]$ using $K_1$ samples, where the corresponding sampling step size is defined as $\delta_{\theta_1}=\pi/(K_1-1)$. For each sampled $\theta_1$, a dedicated nonuniform set of $K_2$ samples is generated for $\theta_M$, following the procedure in \eqref{eq:T} in the Appendix~\ref{app:first}, where $d_\text{R}$ determines the maximum resulting value of $r_{1,1}$ over the constructed intersections. Accordingly, the AoD search space $\boldsymbol{\Theta}$ is characterized by $(d_\text{R}, K_1, K_2)$. In the following experiments, we set $K_1=K_2=360$ (corresponding to $\delta_{\theta_1}\approx 0.501^\circ$), while $d_\text{R}$ is specified separately for each case. For the DL estimation stage, since $\omega=\pi/2$, the orientation search space $\boldsymbol{\Psi}$ reduces to a one-dimensional search over $\varphi\in[0,\pi]$. We uniformly sample this interval using $K_3=360$ points, with step size $\delta_{\varphi}=\pi/(K_3-1)\approx 0.501^\circ$. 
Moreover, unless otherwise specified, perfect quantization is assumed for all exchanged geometric reference parameters, i.e., $q_i=\infty,~i\in \setI_4$.
\begin{figure}
    \centering
    \begin{subfigure}[t]{1\linewidth}
        \centering        \includegraphics[width=\linewidth]{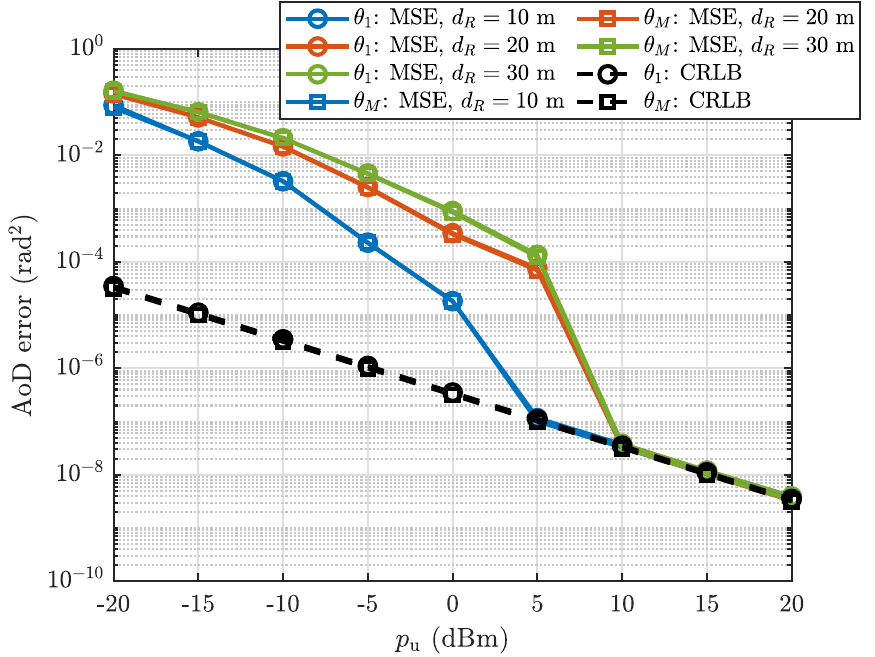}
        \caption{MSE and CRLB of the AoD estimates vs. UE transmit power. 
        } 
        \label{fig:AoD_MSE_CRB_vs_pow}
    \end{subfigure}    
    \begin{subfigure}[t]{1\linewidth}
        \centering       \includegraphics[width=\linewidth]{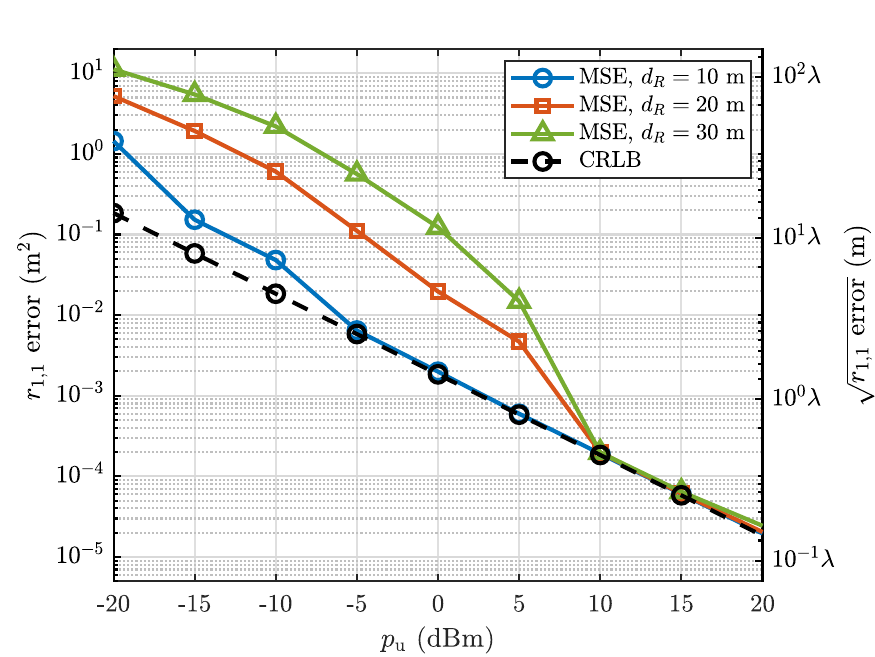}
        \caption{MSE and CRLB of the distance estimate vs. UE transmit power.}
        \label{fig:r_MSE_CRB_vs_pow}
    \end{subfigure}
    \caption{MSE (simulation-based) and CRLB of $(\theta_1,\theta_M,r_{1,1})$ vs. the UE transmit power. } 
    \label{fig:MSE_CRB}
\end{figure}

\begin{figure}[t!]
    \centering
    \includegraphics[width=\linewidth]{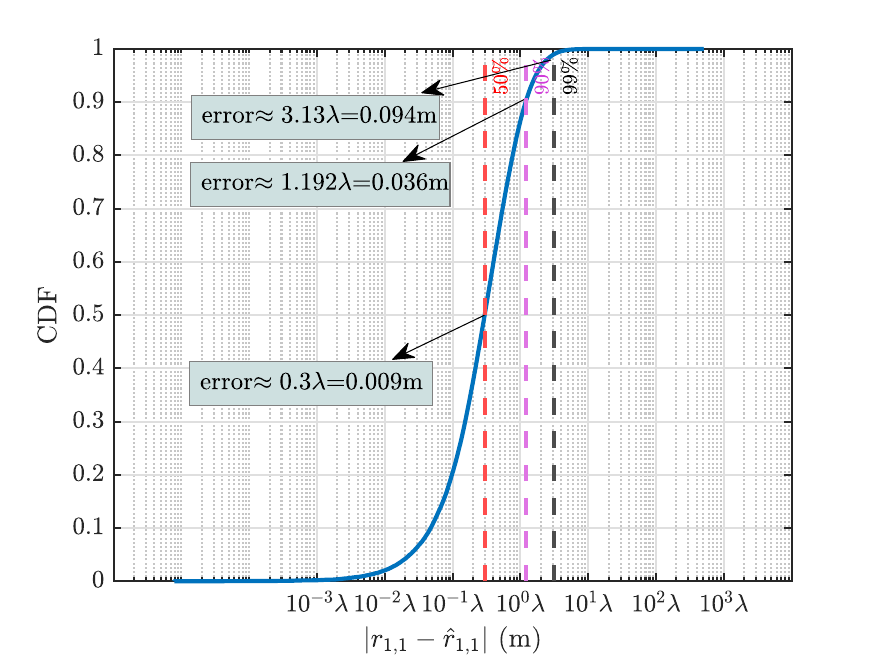}
         \caption{CDF of the first UE antenna distance $r_{1,1}$ estimation error for $p_\text{u}=5\, \rm dBm$ and $d_\text{R}=30\, \mathrm{m}$.}
       \label{fig:CDF_r11_5dbm}
\end{figure}

\begin{figure}
    \centering
    \begin{subfigure}[t]{1\linewidth}
        \centering        \includegraphics[width=\linewidth]{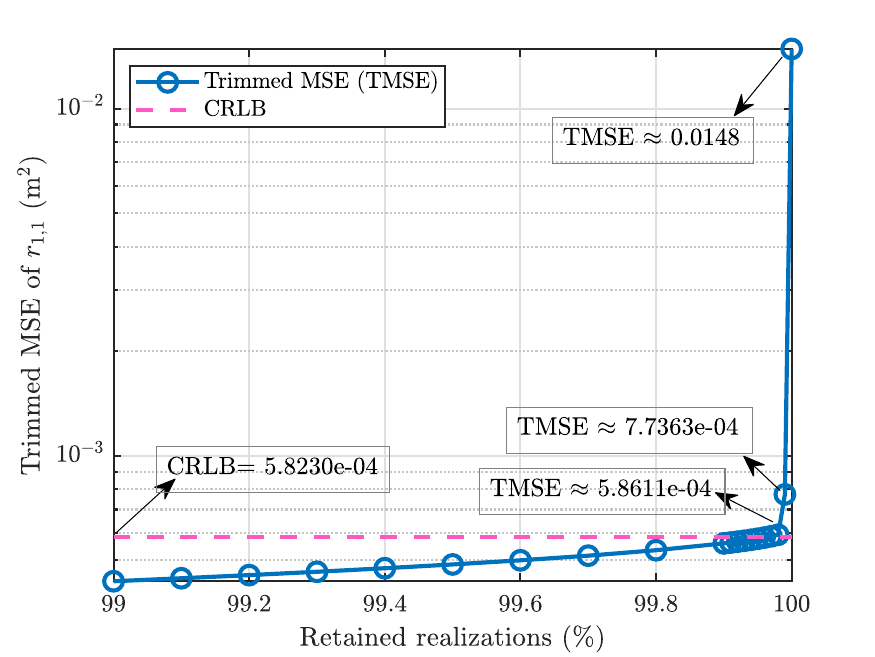}
        \caption{Trimmed MSE and CRLB of $r_{1,1}$ vs. the percentage of retained realizations.
        } 
        \label{fig:TMSE-only_5dbm}
    \end{subfigure}    
    \begin{subfigure}[t]{1\linewidth}
        \centering       \includegraphics[width=\linewidth]{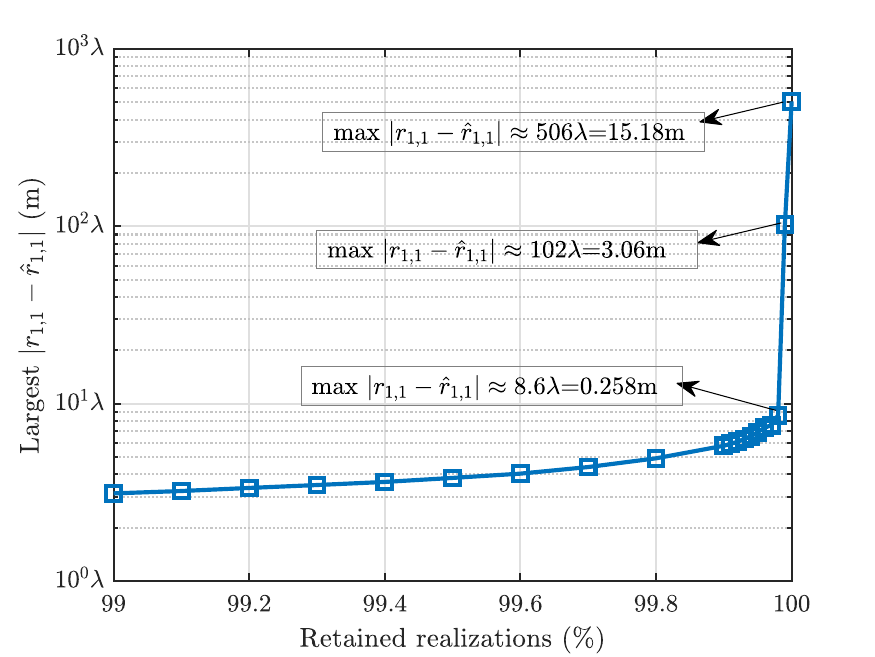}
        \caption{Largest retained $|r_{1,1}-\hat{r}_{1,1}|$ vs. the percentage of retained realizations.}
        \label{fig:max_err_dist_5dbm}
    \end{subfigure}
    \caption{Distance estimation accuracy for $p_\text{u}=5\, \rm dBm$ and $d_\text{R}= 30\, \rm m$, after retaining different percentages of the realizations.}
    \label{fig:TMSE_largest_err_5dbm}
\end{figure}

\begin{figure}[t!]
    \centering
    \includegraphics[width=\linewidth]{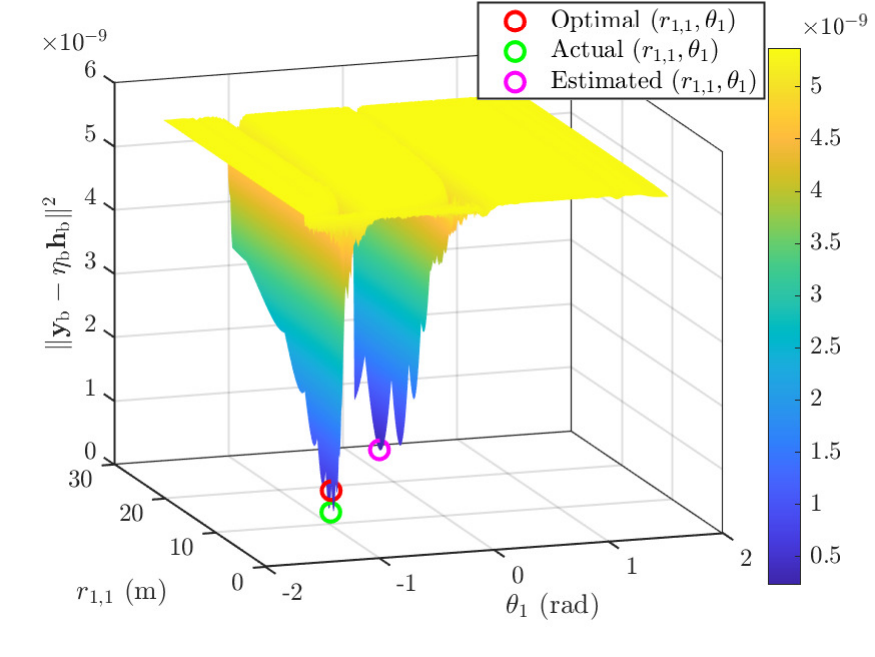}
         \caption {Objective function of the AoD estimator for a representative outlier realization for $p_\text{u}=5\, \rm dBm$ and $d_\text{R}= 30\, \rm m$ case.}
       \label{fig:Likelihood}
\end{figure}

Figs.~\ref{fig:MSE_CRB}-\ref{fig:Likelihood} evaluate the accuracy of the first stage of the proposed LoS MIMO channel parameter estimator. 
%
Fig.~\ref{fig:MSE_CRB} evaluates the accuracy of the reference parameters estimated at the BS from a single UL pilot sequence in terms of the MSE and CRLB for $d_\text{R}\in\{10,20, 30\}\mathrm{m}$. Recall that $d_\text{R}$ is the maximum distance considered by the estimator and can be set according to the available prior range information. A tighter bound, e.g., $d_\text{R}=10~\mathrm{m}$, can be used when the deployment area is known, whereas without such information the search range may be chosen based on the MIMO Rayleigh distance in~\cite{Dai_mixedLoS}, which is $4 l_{\text{b},M}l_{\text{u},N}/\lambda~\approx28 \mathrm{m}$ for the considered setup; hence, $d_\text{R}=30~\mathrm{m}$ is also considered, with $d_\text{R}=20~\mathrm{m}$ as an intermediate case.
%
Fig.~\ref{fig:AoD_MSE_CRB_vs_pow} presents the AoD MSE, i.e., $\Exp\big\{\big| \hat{\theta}_{i}-\theta_{i}\big|^2\big\};~i=1,M$, together with the corresponding CRLB, i.e., $(\mathbf{F}_{\boldsymbol{\theta}}^{-1})_{[1,1]}$ and $(\mathbf{F}_{\boldsymbol{\theta}}^{-1})_{[2,2]}$. 
As shown in Fig.~\ref{fig:AoD_MSE_CRB_vs_pow}, the two AoDs exhibit nearly identical estimation errors. However, the same angular error can result in different distance errors depending on the UE location and the sensitivity of the geometric mapping. Thus, to further assess the reference UE-antenna localization accuracy, Fig.~\ref{fig:r_MSE_CRB_vs_pow} evaluates the estimation error of the distance $r_{1,1}$.
%
From \eqref{eq:r11} the distance estimate is $\hat{r}_{1,1}=r_{1,1}(\hat{\theta}_{1}, \hat{\theta}_{M})$, and its MSE is $\Exp\big\{\big| \hat{r}_{1,1}-r_{1,1}\big|^2\big\}$, while the corresponding CRLB is given by
$\mathbf{g}^\tran \mathbf{F}_{\boldsymbol{\theta}}^{-1}\mathbf{g}$, where $\mathbf{g}=\big[\frac{\partial r_{1,1}}{\partial\theta_1}, \frac{\partial r_{1,1}}{\partial\theta_M}\big]^\tran$ and the proof is in {Appendix~\ref{app:r_11}}.
%
Note that, to mitigate the effects of the finite grid resolution of the search space $\boldsymbol{\Theta}$, a local refinement has been applied to all points using a finer search grid centered at the initial estimate. 

Fig.~\ref{fig:MSE_CRB} shows that at sufficiently high $p_\text{u}$, the MSEs of both the AoD and reference-distance estimates approach the corresponding CRLBs, indicating a local estimation regime in which the estimator operates around the correct minimum of the objective function in~\eqref{eq:th1_thM_hat}, with small residual estimation errors.\footnote{
Exploiting prior information restricting $\theta_1$, e.g., 
to $(-\pi/3,\pi/3)$ can further improve the AoD and distance estimation accuracy.}.
As $p_\text{u}$ decreases, particularly for larger $d_\text{R}$, the MSE deviates sharply from the CRLB, indicating threshold behavior of the estimator, in which noise-induced ambiguities between competing minima of the estimator cost function occasionally lead to large estimation errors. Accordingly, increasing $d_\text{R}$ shifts the transition to the local regime toward higher transmit powers.
In addition, the local refinement is effective for $d_\text{R}=10\mathrm{m}$ from approximately $p_\text{u}=0\mathrm{dBm}$, whereas for $d_\text{R}=20\mathrm{m}$ and $d_\text{R}=30\mathrm{m}$, similar behavior is observed from approximately $p_\text{u}=5\mathrm{dBm}$. Beyond these power levels, the MSE curves become nearly indistinguishable from the corresponding CRLBs. In the threshold regime, refinement provides limited improvement when the initial estimate is near an incorrect local minimum.

%
To further characterize the threshold behavior, Figs.~\ref{fig:CDF_r11_5dbm}–\ref{fig:Likelihood} consider $p_\text{u}=5\:\mathrm{dBm}$ and $d_\text{R}=30\:\mathrm{m}$.  Fig.~\ref{fig:CDF_r11_5dbm} presents the cumulative distribution function (CDF) of the reference-distance estimation error and shows that the vast majority of realizations achieve high accuracy.
The median, $90^\text{th}$-percentile, and $99^\text{th}$-percentile errors are approximately $0.009\,\rm m$, $0.036\,\rm m$, and $0.094\,\rm m$, respectively. Thus, for most noise realizations, the estimator remains close to the correct solution and yields highly accurate distance estimates.
This behavior is further illustrated by the trimmed-MSE results in Fig.\ref{fig:TMSE-only_5dbm}. The trimmed MSE is obtained by sorting the squared distance-estimation errors and averaging only a specified fraction of the realizations with the smallest errors. 
This allows us to examine the contribution of rare large errors to the overall MSE. When approximately $99.9\%$ of the realizations are retained, the trimmed MSE is $5.86\times10^{-4},\mathrm{m}^2$, close to the conventional CRLB of $5.823\times10^{-4},\mathrm{m}^2$, which is included as a reference for local estimation accuracy\footnote{Note that the conventional CRLB is not a lower bound on the trimmed MSE, as the retained realizations are selected based on their estimation errors.}. 
Including the remaining realizations causes a sharp increase in the MSE to approximately $1.48\times10^{-4},\mathrm{m}^2$, demonstrating the significant contribution of rare large errors to the overall estimation performance.
Correspondingly, Fig.~\ref{fig:max_err_dist_5dbm} shows that the largest distance error among the retained $99.9\%$ of the realizations is approximately $0.26,\mathrm{m}$, whereas the excluded $0.1\%$ contains errors of several meters, reaching approximately $15,\mathrm{m}$. Together with Fig.\ref{fig:TMSE-only_5dbm} these results illustrate the threshold behavior: in the local-error regime, most estimates remain near the correct solution, yielding a trimmed MSE close to the CRLB, whereas rare nonlocal errors associated with competing minima of the estimator objective function can dominate the overall MSE.
%
The large error values are typically concentrated near the angular boundaries and at relatively large distances.
Fig. \ref{fig:Likelihood} further illustrates this behavior through the estimator objective function for a representative UE geometry exhibiting a large estimation error. Due to geometric ambiguities, distinct UE locations may produce similar spatial channel responses across the finite BS aperture, giving rise to competing local minima at substantially different angle and distance values. In the presence of noise, an incorrect minimum may attain a lower cost than the minimum near the true parameters, resulting in a large estimation error not captured by the local CRLB. Such errors can be mitigated by increasing the UL pilot power or exploiting prior distance information to restrict the estimation search space.
Overall, Figs.~\ref{fig:MSE_CRB}-\ref{fig:Likelihood} show that a single UL pilot sequence can provide sufficiently accurate reference-angle and distance estimates for the subsequent DL orientation-estimation stage, depending on the operating conditions.
%
\begin{figure}[t!]
    \centering
    \includegraphics[width=\linewidth]{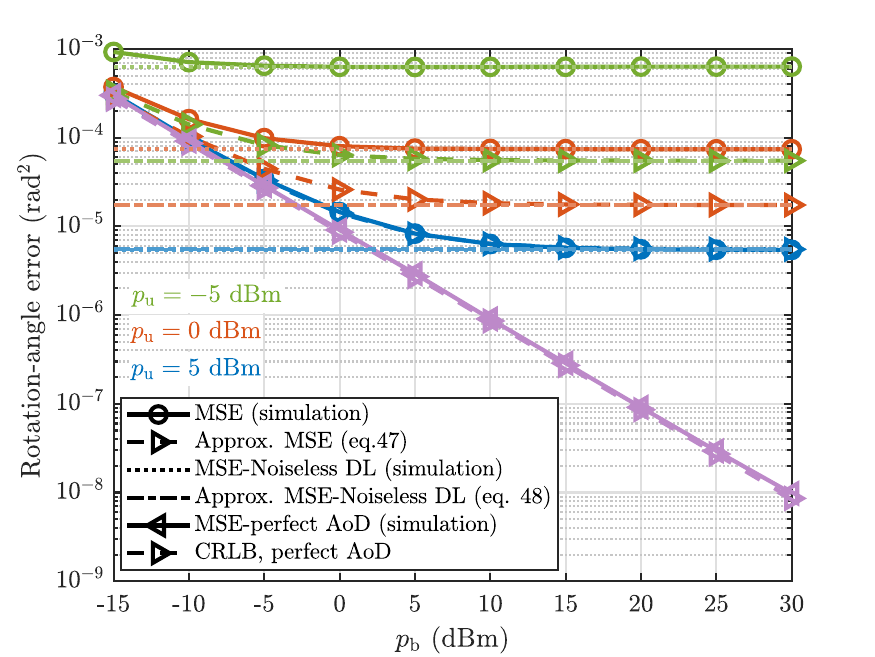}
         \caption{Rotation-angle estimation error vs. BS  power $p_\text{b}$
 for different UE powers $p_\text{u}$, and assuming $d_\text{R}= 10\, \rm m$ in the UL estimation.}
       \label{fig:rot_error}
\end{figure}
%

Fig.\ref{fig:rot_error} evaluates the rotation-angle estimation accuracy in terms of the MSE, $\mathbb{E}\{|\hat{\varphi}-\varphi|^2$, and its theoretical approximations. In all simulated cases, a few rounds of local refinement are applied around the initial estimate, using progressively smaller step sizes than $\delta_\varphi$.
For unbiased estimates of $\boldsymbol{\theta}$ and $\varphi$, the first-order MSE approximation of $\hat{\varphi}$ is obtained from \eqref{eq:Cov_e_Final_theta_hat} by setting $F_\varphi=(\mathbf{F}_{\boldsymbol{\nu}})_{[1,1]}$ and $\mathbf{s}_{{\varphi}\boldsymbol{\theta}}=(\mathbf{S}_{\boldsymbol{\nu}\boldsymbol{\theta}})_{[1,:]}$, yielding
%
%
\begin{align}\label{eq:MSE_phi}
\mathbb{E}\{|\hat{\varphi}-\varphi|^2\}\approx F_\varphi^{-1}+F_\varphi^{-2}{\mathbf{s}}_{{\varphi}\boldsymbol{\theta}}{\mathbf{F}}_{\boldsymbol{\theta}}^{-1}{\mathbf{s}}^\tran_{{\varphi}\boldsymbol{\theta}}.
\end{align}
In the absence of DL observation noise ($\sigma_\text{u}=0$), only the propagated first-stage estimation error remains. The MSE of $\varphi$ is then approximated by
\begin{align}\label{eq:MSE_phi_DL_noiseless}
\mathbb{E}\{|\hat{\varphi}_{\sigma_\text{u}=0}-\varphi|^2\}\approx \lim_{\sigma_\text{u}\rightarrow0}\ F_\varphi^{-2}{\mathbf{s}}_{{\varphi}\boldsymbol{\theta}}{\mathbf{F}}_{\boldsymbol{\theta}}^{-1}{\mathbf{s}}^\tran_{{\varphi}\boldsymbol{\theta}}.
\end{align}
%
%
%
%
%
%
Fig.\ref{fig:rot_error} shows that 
with perfect AoD knowledge, the simulated MSE closely follows the corresponding CRLB, $F_\varphi^{-1}$, and decreases continuously with $p_\text{b}$, indicating accurate DL estimation in the absence of first-stage uncertainty. 
%
When estimated AoDs are used, the approximations in \eqref{eq:MSE_phi} and \eqref{eq:MSE_phi_DL_noiseless} closely agree with the corresponding simulated MSEs when the first-stage estimation errors are sufficiently small (i.e., in the local-error regime), particularly for $p_\text{u}=5 \rm \:dB$. As $p_\text{u}$ decreases, the gap between the approximations and simulated MSEs increases due to the greater occurrence of nonlocal AoD estimation errors. This is consistent with the threshold behavior observed in Figures~\ref{fig:CDF_r11_5dbm}–\ref{fig:Likelihood}: the CRLB-based first-order analysis characterizes errors near the correct solution but becomes optimistic when the estimator occasionally selects an incorrect minimum of the first-stage objective function. Such errors can propagate into the rotation-angle estimate and are not fully captured by the local approximation.
%
Another important observation is that the rotation-angle MSE and its approximation in \eqref{eq:MSE_phi} initially decrease with increasing $p_\text{b}$. At sufficiently high $p_\text{b}$, however, the DL observation noise becomes negligible, and the estimation performance approaches the corresponding noiseless-DL ($\sigma_\text{u}=0$) case. Further increasing $p_\text{b}$ therefore provides little improvement, as the remaining error is dominated by the propagation of first-stage AoD estimation errors, resulting in an error floor.
Overall, these results show that two DL pilot sequences can accurately estimate the remaining UE-array orientation when sufficiently accurate reference-angle estimates are available from the UL stage.
%

\begin{figure}
    \centering
    \begin{subfigure}[t]{1\linewidth}
        \centering        \includegraphics[width=\linewidth]{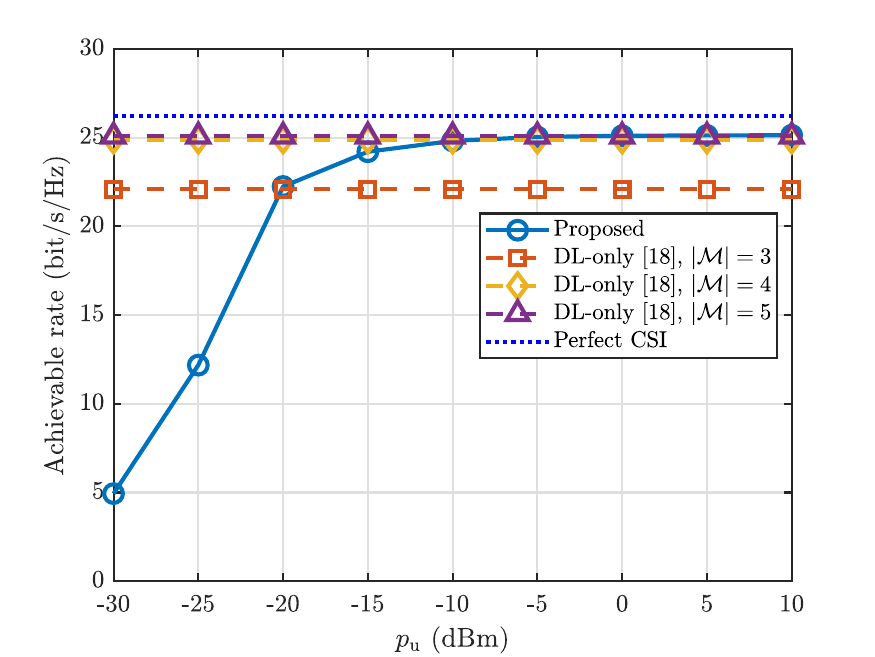}
        \caption{Rate versus UE transmit power: proposed scheme vs.~\cite{Shima}.}
        \label{fig:Rate_compare}
    \end{subfigure}    
    \begin{subfigure}[t]{1\linewidth}
        \centering       \includegraphics[width=\linewidth]{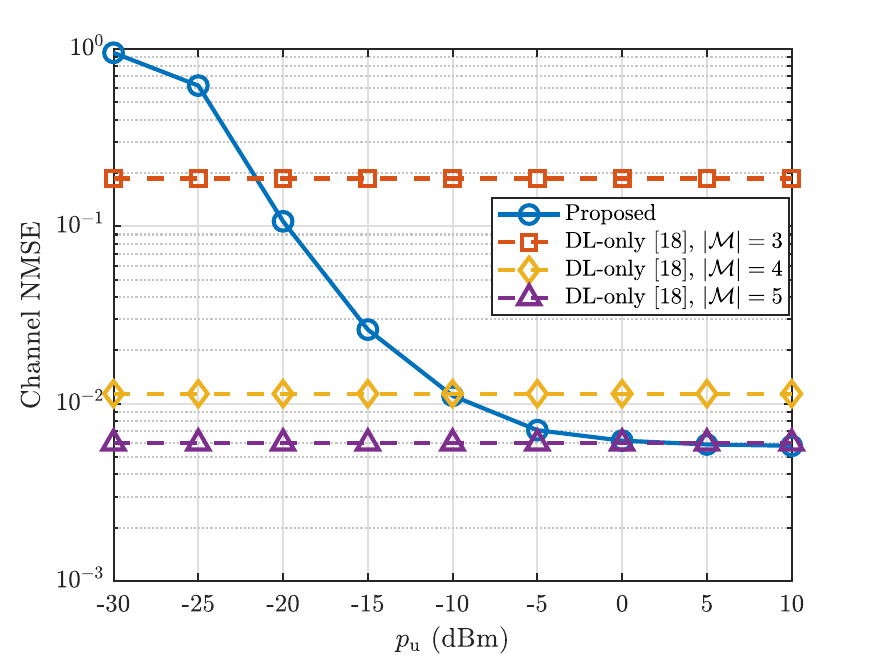}
        \caption{Channel NMSE versus UE transmit power: proposed scheme vs.~\cite{Shima}.}
        \label{fig:NMSE_compare}
    \end{subfigure}
    \caption{Channel estimation accuracy of the proposed method versus the DL-pilot-only scheme in~\cite{Shima}.} 
    \label{fig:Rate_NMSE_compare}
\end{figure}

\begin{figure}[t!]
    \centering
    \includegraphics[width=\linewidth]{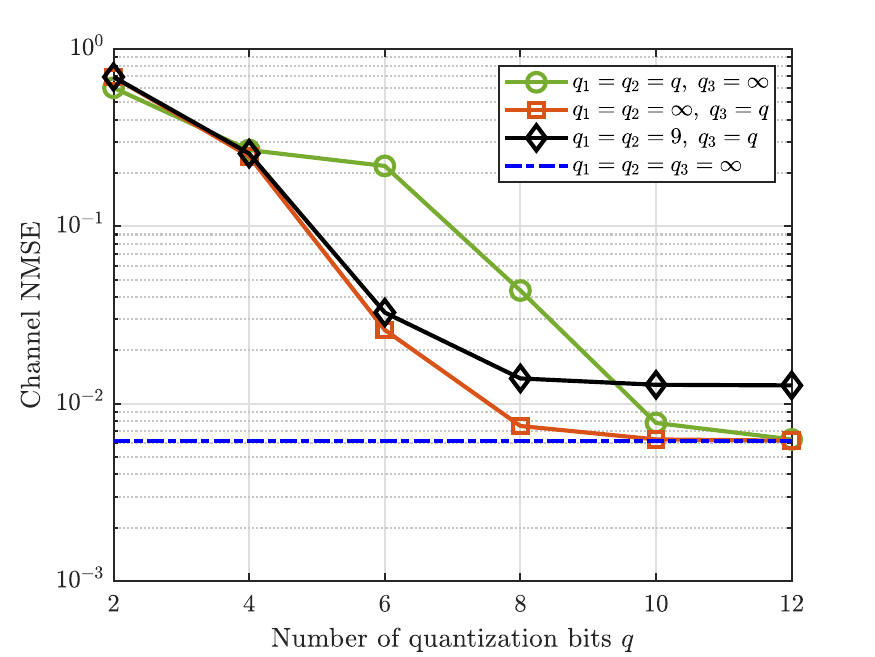}
         \caption{Channel NMSE versus angular quantization level.}
       \label{fig:NMSE_q}
\end{figure}

\color{black}
Now we evaluate the accuracy of the reconstructed LoS MIMO channel using the achievable rate and the normalized MSE (NMSE) in Figs.~\ref{fig:Rate_NMSE_compare} and~\ref{fig:NMSE_q}. The results are obtained with $p_\text{b}=0\:\mathrm{dBm}$, $d_\text{R}=10\: \mathrm{m}$, 
while $p_\text{u}$ is specified for each case. The reported results do not include local angular refinement; incorporating such refinement can further improve the estimation accuracy, particularly at moderate and high SNRs. 
We define the UE rate as 
\begin{align}
\label{eq:rate}
R\triangleq\sum_{s=1}^S\log_2\Big(1+\frac{p_s|\mathbf{u}_s^\herm\mathbf{H}\hat{\mathbf{m}}_s|^2}{\sum_{\bar{s}\neq s}^Sp_{\bar{s}}|\mathbf{u}_s^\herm\mathbf{H}\hat{\mathbf{m}}_{\bar{s}}|^2+\|\mathbf{u}_s\|^2 \sigma_\text{u}^2}\Big),
\end{align}
where, $S$ is the number of data streams (number of strongest singular values of $\hat{\mathbf{H}}$), $p_s$ is the power allocated to the $s^\text{th}$ stream using water-filling over $\hat{\mathbf{H}}$ such that $\sum_s p_s\leq P$, with $P=-5~\text{dBm}$ after conversion to linear units.
Moreover, 
$\hat{\mathbf{m}}_s$ is the $s^\text{th}$ right singular vector of $\hat{\mathbf{H}}$ as the TX beamformer, and $\mathbf{u}_s=(\sum_s p_s\mathbf{H}\hat{\mathbf{m}}_s\hat{\mathbf{m}}_s^\herm\mathbf{H}^\herm+\sigma_\text{u}^2\mathbf{I})^{-1}\mathbf{H}\hat{\mathbf{m}}_s$ is the MMSE receiver. 
The NMSE defined as $\|\mathbf{H}-\hat{\beta}\hat{\mathbf{H}}\|_\text{F}^2/\|\mathbf{H}\|_\text{F}^2$, where we adopt $\hat{\beta}$ 
to avoid the impact of scalar gain ambiguity, i.e.,
\begin{align}
\label{eq:beta}
\hat{\beta}&= \mathop{\text{argmin}}_{\substack{\beta\in \mathbb{C}}} \|\mathbf{H}-\beta\hat{\mathbf{H}}\|_\text{F}^2= \frac{\text{tr}(\mathbf{H}\hat{\mathbf{H}}^\herm)}{\text{tr}(\hat{\mathbf{H}}\hat{\mathbf{H}}^\herm)}.
\end{align}

%
Fig.~\ref{fig:Rate_NMSE_compare} 
compares the proposed scheme, using one UL and two DL pilot sequences, with the DL-only scheme in~\cite{Shima} using three, four, or five DL pilot sequences, i.e., $|\setM|=\{3,4,5\}$.
For the DL-only baseline, we adopt the nonuniform pilot-antenna patterns reported in~\cite{Shima}, where the two outermost BS antennas remain active, and the interior antennas are selected such that
$\setM=\{1,3,64\}$, $\{1,4,60,64\}$, and $\{1,4,9,64\}$ for $|\setM|=3$, $4$, and $5$, respectively.
%
Since the total BS transmit power $p_\text{b}$ is equally divided among the $|\setM|$ pilot-transmitting antennas, we set $\tau_\text{b}=2|\setM|$, such that the effective training energy per active antenna, $\tau_\text{b}p_\text{b}/|\setM|=2p_\text{b}$, is the same as in the proposed scheme with $|\setM|=2$ and $\tau_\text{b}=4$.
Table~\ref{tab:overhead_comparison} shows the corresponding pilot transmission, feedback, and UE-side search requirements. The proposed scheme shifts part of the acquisition to the UL stage, thereby reducing the DL pilot requirement and UE-side search dimension, whereas the DL-only scheme performs the complete geometric-parameter estimation at the UE. The two approaches can therefore be suitable for different operating conditions; for example, the DL-only scheme can be attractive when UL pilot transmission is unavailable or highly constrained.
%
%

It is observed from Fig.~\ref{fig:Rate_NMSE_compare}  that the proposed scheme, using one UL and two DL pilot sequences, achieves both achievable rate and channel NMSE comparable to the four-pilot DL-only scheme already at $p_\text{u}=-10\:\mathrm{dBm}$. As $p_\text{u}$ increases, its performance approaches that of the five-pilot DL-only scheme, with nearly identical rate and NMSE around $p_\text{u}=0\:\mathrm{dBm}$. Thus, by exploiting a single UL pilot, the proposed scheme can achieve comparable end-to-end performance while requiring only two DL pilot sequences.
The differences among the schemes are more pronounced in NMSE than in achievable rate, since NMSE directly reflects the channel reconstruction error, whereas the rate is less sensitive to some channel mismatches in the considered single-UE setting. Therefore, we use NMSE to further examine the impact of parameter quantization.
\begin{table*}[t]
\centering
\caption{Pilot transmission and feedback requirements of the considered channel acquisition schemes.}
\label{tab:overhead_comparison}
\renewcommand{\arraystretch}{1.15}
\setlength{\tabcolsep}{5pt}
\scriptsize
\resizebox{\textwidth}{!}{%
\begin{tabular}{c|c|c|c|c|c|c|c}
\hline
\textbf{Scheme}
&
\textbf{UL stage}
&
\textbf{DL stage}
&
\textbf{DL energy/antenna}
&
\textbf{Total DL energy}
&
\textbf{DL-to-UE feedback}
&
\textbf{UE-to-BS feedback}
&
\textbf{UE search dimension}
\\
\hline
DL-only~\cite{Shima}
& -- 
& $|\setM|=3,\ \tau_{\mathrm b}=6$
& $2p_{\mathrm b}$
& $6p_{\mathrm b}$
& --
& Two AoDs and UE orientation
& 3-D/4-D
\\

DL-only~\cite{Shima}
& --
& $|\setM|=4,\ \tau_{\mathrm b}=8$
& $2p_{\mathrm b}$
& $8p_{\mathrm b}$
& --
& Two AoDs and UE orientation
& 3-D/4-D
\\

DL-only~\cite{Shima}
& --
& $|\setM|=5,\ \tau_{\mathrm b}=10$
& $2p_{\mathrm b}$
& $10p_{\mathrm b}$
& --
& Two AoDs and UE orientation
& 3-D/4-D
\\
\hline

Proposed
& $\tau_{\mathrm u}=1$
& $|\setM|=2,\ \tau_{\mathrm b}=4$
& $2p_{\mathrm b}$
& $4p_{\mathrm b}$
& Two AoDs
& UE orientation
& 1-D/2-D
\\
\hline
\end{tabular}%
}

\end{table*}
Fig.~\ref{fig:NMSE_q} examines the impact of angular quantization on the channel NMSE of the proposed scheme at $p_\text{u}=p_\text{b}=0\: \mathrm{dBm}$. Here, $(q_1, q_2,q_3)$ denote the numbers of quantization bits for $(\theta_1,\theta_M,\varphi)$, respectively. Accordingly, the reconstructed channel at the BS is obtained as $\hat{\mathbf{H}}(\hat{\theta}_1,\hat{\theta}_M, \tilde{\varphi})$, where $\tilde{\varphi}=Q_{q_3}^{-1} \big( Q_{q_3}(\hat{\varphi})\big)$. Moreover, $\hat{\varphi}$ is obtained at the UE using the de-quantized AoDs $(\tilde{\theta}_1,\tilde{\theta}_M)$ defined in \eqref{eq:AoD_tilde}. It is observed that when only $\hat{\varphi}$ is quantized, i.e., $q_1=q_2=\infty$, the NMSE approaches the perfect-quantization performance with about $8$ bits. In contrast, when only the two AoD estimates are quantized, i.e., $q_3=\infty$, about 10--12 bits per AoD are required to approach the same level. Moreover, for $q_1=q_2=9$, increasing $q_3$ beyond about 8 bits provides little additional improvement, since the residual error is then mainly limited by the AoD quantization. These results indicate that a finer quantization resolution is more important for the AoDs than for the UE orientation.
%
\color{black}
%
%
 
\section{Conclusions}
\label{sec:CONC}
A two-stage hybrid UL/DL geometry-based N-F LoS MIMO channel acquisition scheme is proposed for asymmetric UE transmit/receive configurations with a single UL-active antenna. The BS first estimates the reference-location geometry from one UL pilot sequence, while the UE estimates the remaining array orientation from two DL pilot sequences. By exploiting the geometric information shared between the UL and DL, the proposed strategy flexibly divides the acquisition burden between the BS and UE without requiring pilot transmission from all UE antenna elements. We derived CRLBs and a first-order analysis that quantifies the propagation of first-stage uncertainty into the orientation estimate. Numerical results characterized the estimator threshold behavior and demonstrated an orientation-estimation error floor at high DL SNR due to propagated first-stage errors, while showing accurate end-to-end channel reconstruction. Since the approach relies on shared physical geometry rather than complex-channel reciprocity, it is applicable to both TDD and FDD operation.
\appendices

\section{AoD search space construction}
\label{app:first}
Here, we describe the sampling strategy used to construct the angular search space $\boldsymbol{\Theta}$. Similar to \cite{Dai_mixedLoS, Localization}, the first AoD is sampled uniformly over the feasible angular range as 
\begin{equation}
\label{eq:U}
\begin{aligned}
  \setU=\Big\{ \theta_\text{min}+(k_1-1)\frac{\theta_\text{max}-\theta_\text{min}}{K_1-1}\quad \forall k_1\in \setI_{K_1}\Big\},   
\end{aligned}
\end{equation}
where $K_1$ denotes the number of samples.
In the N-F, uniformly sampling both $\theta_1$ and $\theta_M$ does not result in a desirable distribution of the UE location points. Inspired by the sampling strategy of \cite{Localization}, we therefore sample $\theta_M$ such that, together with a given $\theta_1$, the resulting intersection points span the distance range in a non-uniform manner. More specifically, the generated UE locations are intended to be denser at shorter distances from the BS and progressively sparser at larger distances.
To achieve this, for each $\theta_1\in\setU$, a dedicated $\theta_M$ sample set, $\setT_{\theta_1}$ is generated.
To obtain the desired distribution of UE locations, we first consider the exponential distance grid with $K_2$ samples as proposed in \cite{Localization}, i.e., $\{d_\text{R}^\frac{k_2}{K_2},
~
k_2\in\setI_{K_2},\}$, which spans the distance range $[1,d_\text{R}]$ with finer resolution at shorter distances and coarser resolution at larger distances. Using the angle-distance mapping relation in \eqref{eq:r11}, the corresponding samples of $\theta_M$ are determined as
\begin{equation}
\label{eq:T}
\begin{aligned}
   \setT_{\theta_1}=\Bigg\{\tan^{-1}\Bigg( \frac{l_{b,M}+d_\text{R}^\frac{k_2}{K_2}\sin\theta_1}{d_\text{R}^\frac{k_2}{K_2}\cos\theta_1}\Bigg); ~ \forall k_2\in \setI_{K_2}, \theta_1\in \setU\Bigg\}, 
\end{aligned}
\end{equation}
The parameter $d_\text{R}$ can be set to the Rayleigh distance or the maximum expected user distance. The resulting joint angular search space is given by
\begin{equation}
\label{eq:Theta_grid}
\boldsymbol{\Theta}
=
\left\{
(\theta_1,\theta_M)
:\;
\theta_1\in\setU,\;
\theta_M\in\setT_{\theta_1}
\right\}.
\end{equation}
\section{Derivation of FIM for AoDs}
\label{app:A1}
Let us define $\boldsymbol{\rho}\triangleq[ \Re{(\eta_\text{b})}, \Im (\eta_\text{b}), \theta_1, \theta_M,]^\tran$ as the real-valued reference parameter vector of the UL estimation phase. Thus, the $(i,j)^\text{th}$ element of the full FIM, i.e.,  $\mathbf{F}_{\boldsymbol{\rho}}=\mathbb{C}^{4\times4}$ is given as~\cite{CRLB}
\begin{equation}
\label{eq:F_full}   (\mathbf{F}_{\boldsymbol{\rho}})_{[i,j]} = \frac{2\tau_\text{u}} {\sigma^2_\text{b}}\Re\Big(\frac{\partial\eta^*_{\text{b}}\mathbf{h}^\herm_{\text{b},1}}{\partial \boldsymbol{\rho}_{[i]}}\frac{\partial\eta_{\text{b}}\mathbf{h}_{\text{b},1}}{\partial \boldsymbol{\rho}_{[j]}}\Big),~\: i,j\in\setI_4
\end{equation}
where $\mathbf{h}_{\text{b},1}=\mathbf{h}_{\text{b},1}(\theta_1, \theta_M)$, with the arguments omitted for notational simplicity. Let $\boldsymbol{\eta}_{\text{b}}\triangleq\boldsymbol[\Re{(\eta_\text{b})}, \Im (\eta_\text{b})]^\tran={\rho}_{[1:2]}$ denote the gain vector, while $\boldsymbol{\theta}=[\theta_1, \theta_M ]^\tran=\boldsymbol{\rho}_{[3:4]}$ is the previously defined AoD vector.
Thus, the FIM $\mathbf{F}_{\boldsymbol{\rho}}$ can then be partitioned into gain-related and AoD-related blocks as
\begin{equation}
\label{eq:F_full2}   \mathbf{F}_{\boldsymbol{\rho}} = \begin{bmatrix}
\tilde{\mathbf{F}}_{\boldsymbol{\eta}_\text{b}} & \tilde{\mathbf{F}}_{\boldsymbol{\eta}_\text{b}\boldsymbol{\theta}}  \\
\tilde{\mathbf{F}}_{\boldsymbol{\theta}\boldsymbol{\eta}_\text{b}} & \tilde{\mathbf{F}}_{\boldsymbol{\theta}}  \\
\end{bmatrix},
\end{equation}
where 
\begin{equation}
\label{eq:F_eta}   \tilde{\mathbf{F}}_{\boldsymbol{\eta}_\text{b}} = (\mathbf{F}_{\boldsymbol{\rho}})_{[1:2;1:2]}=\frac{2\tau_\text{u}} {\sigma^2_\text{b}}\| \mathbf{h}_{\text{b},1}\|^2\mathbf{I}_2,
\end{equation}
\begin{equation}
\label{eq:F_thet}   \tilde{\mathbf{F}}_{\boldsymbol{\theta}} = (\mathbf{F}_{\boldsymbol{\rho}})_{[3:4;3:4]}=\frac{2\tau_\text{u}|\eta_\text{b}|^2} {\sigma^2_\text{b}}\Re\Big( \mathbf{J}_{\boldsymbol{\theta}}^\herm\mathbf{J}_{\boldsymbol{\theta}}\Big),
\end{equation}
where $\mathbf{J}_\text{b}$ is defined as in~\eqref{eq:J_b}.
Moreover, for cross terms we have $\tilde{\mathbf{F}}_{\boldsymbol{\eta}_\text{b}\boldsymbol{\theta}}=\tilde{\mathbf{F}}_{\boldsymbol{\theta}\boldsymbol{\eta}_\text{b}}^\tran$ where
\begin{equation}
\label{eq:F_nu_thet}   \tilde{\mathbf{F}}_{\boldsymbol{\eta}_\text{b}\boldsymbol{\theta}} = (\mathbf{F}_{\boldsymbol{\rho}})_{[1:2;3:4]}=\frac{2\tau_\text{u}} {\sigma^2_\text{b}}\Re\Big(\begin{bmatrix}
1 ,~-\mathrm{j}
\end{bmatrix}^\tran\eta_{\text{b},1}\mathbf{h}^\herm_{\text{b},1}\mathbf{J}_\text{b}\Big).
\end{equation}
Using the Schur complement, the FIM of $\boldsymbol{\theta}$ accounting for the unknown gain parameter $\eta_\text{b}$ is given by
\begin{equation}
\label{eq:F_AoD_SC}
\begin{aligned}   \mathbf{F}_{\boldsymbol{\theta}}&= \tilde{\mathbf{F}}_{\boldsymbol{\theta}}-\tilde{\mathbf{F}}_{\boldsymbol{\theta}\boldsymbol{\eta}_\text{b}}\tilde{\mathbf{F}}_{\boldsymbol{\eta}_\text{b}}^{-1}\tilde{\mathbf{F}}_{\boldsymbol{\eta}_\text{b}\boldsymbol{\theta}} \\ &=\frac{2\tau_\text{u}} {\sigma^2_\text{b}}\Re\Big(|\eta_\text{b}|^2\mathbf{J}_\text{b}^\herm\mathbf{J}_\text{b}-\|\mathbf{h}_{\text{b},1}\|^{-2}\mathbf{J}_\text{b}^\herm\eta_\text{b}^*\mathbf{h}_{\text{b},1}\mathbf{h}_{\text{b},1}^\herm\eta_\text{b}\mathbf{J}_\text{b}\Big). 
\end{aligned}
\end{equation}
After simplification, we obtain
\begin{equation}
\label{eq:F_AoD_final}
\begin{aligned}   \mathbf{F}_{\boldsymbol{\theta}}=\frac{2\tau_\text{u}|\eta_\text{b}|^2} {\sigma^2_\text{b}}\Re\Big(\mathbf{J}_\text{b}^\herm\underbrace{\big(\mathbf{I}_M-\frac{\mathbf{h}_{\text{b},1}\mathbf{h}_{\text{b},1}^\herm}{\|\mathbf{h}_{\text{b},1}\|^{2}}\big)}_{\triangleq \mathbf{P}_\text{b}} \mathbf{J}_\text{b} \Big),
\end{aligned}
\end{equation}
where $\mathbf{P}_\text{b}$ is the projection matrix onto the subspace orthogonal to $\mathbf{h}_{\text{b},1}$. It is observed that 
the effect of the unknown gain $\eta_\text{b}$ can be incorporated by first deriving the standard FIM expression for $\boldsymbol{\theta}$ assuming known $\eta_\text{b}$, i.e., $2\tau_\text{u}|\eta_\text{b}|^2\sigma^{-2}_\text{b} \Re(\mathbf{J}_\text{b}^\herm\mathbf{J}_\text{b})$ and then replacing the channel derivative matrix $\mathbf{J}_\text{b}$ by its orthogonal projection $\mathbf{P}_\text{b}\mathbf{J}_\text{b}$.
We use this interpretation in~\ref{DL_error}.
\section{CRLB of reference distance \texorpdfstring{$r_{1,1}$}{r11}}
\label{app:r_11}
To obtain the CRLB of $r_{1,1}$, we exploit the deterministic relationship between
$r_{1,1}$ and
$(\theta_1,\theta_M)$, i.e., $\mathbf{h}_\text{b}(\theta_1,\theta_M)=\mathbf{h}_\text{b}(\theta_1, r_{1,1}(\theta_1,\theta_M))$. Accordingly, the derivative terms in $\mathbf{J}_{\boldsymbol{\theta}}$ defined in \eqref{eq:J_b} can be rewritten using the chain rule as
\begin{equation}
\label{eq:partial_h_new}
\begin{aligned}
\frac{\partial\mathbf{h}_\text{b}}{\partial \theta_1}&=\frac{\partial\mathbf{h}_\text{b}(\theta_1,.)}{\partial \theta_1}+\frac{\partial\mathbf{h}_\text{b}(., r_{1,1})}{\partial r_{1,1}}\frac{\partial r_{1,1}(\theta_1,.)}{\partial \theta_1},\\
\frac{\partial\mathbf{h}_\text{b}}{\partial \theta_M}&=\frac{\partial\mathbf{h}_\text{b}(., r_{1,1})}{\partial r_{1,1}}\frac{\partial r_{1,1}(.,\theta_M)}{\partial \theta_M},
\end{aligned}
\end{equation}
where the unspecified argument is held constant during partial differentiation.
Next, we introduce the equivalent parameter vector $\boldsymbol{\alpha}\triangleq\big[\theta_1, r_{1,1}]^\tran$, with the corresponding derivative matrix  $\mathbf{J}_{\boldsymbol{\alpha}}\triangleq\big[\frac{\partial\mathbf{h}_\text{b}}{\partial \theta_1},\frac{\partial\mathbf{h}_\text{b}}{\partial r_{1,1}} \big]$. 
Substituting the above derivatives into \eqref{eq:J_b} yields, $\mathbf{J}_{\boldsymbol{\theta}}=\mathbf{J}_{\boldsymbol{\alpha}}\mathbf{G}$, where
\begin{equation}
\label{eq:G}
\begin{aligned}
\mathbf{G}=\begin{bmatrix} 1 & 0\\ \frac{\partial r_{1,1}}{\partial \theta_1}& \frac{\partial r_{1,1}}{\partial \theta_M}\end{bmatrix}.
\end{aligned}
\end{equation}
Thus, $\mathbf{F}_{\boldsymbol{\theta}}$ in \eqref{eq:F_nub} is rewritten as 
\begin{equation}
\label{eq:F_theta_represent}   \mathbf{F}_{\boldsymbol{\theta}} = \mathbf{G}^\tran\Bigg(\frac{2\tau_\text{u} |\eta_\text{b}|^2} {\sigma^2_\text{b}}\Re\Big(\mathbf{J}_{\boldsymbol{\alpha}}^\herm\mathbf{P}_\text{b}\mathbf{J}_{\boldsymbol{\alpha}}\Big)\Bigg)\mathbf{G}=\mathbf{G}^\tran \mathbf{F}_{\boldsymbol{\alpha}}\mathbf{G} ,
\end{equation}
where $\mathbf{F}_{\boldsymbol{\alpha}}$ denotes the FIM associated with the parameter vector $\boldsymbol{\alpha}$. 
Consequently, $\mathbf{F}_{\boldsymbol{\alpha}}^{-1}=\mathbf{G} \mathbf{F}_{\boldsymbol{\theta}}^{-1}\mathbf{G}^\tran$, and the CRLB of $r_{1,1}$ is given by $(\mathbf{F}_{\boldsymbol{\alpha}}^{-1})_{[2,2]}$. Defining $\mathbf{g}^\tran\triangleq \mathbf{G}_{[2,:]}$, the CRLB of $r_{1,1}$ can be compactly expressed as $\mathbf{g}^\tran \mathbf{F}_{\boldsymbol{\theta}}^{-1}\mathbf{g}$.

\bibliographystyle{IEEEtran}
\let\v\originalcaron
\bibliography{IEEEabbr,refs}

\end{document}